\documentclass[preprint,tightenlines]{elsarticle}
\usepackage{amsmath}
\usepackage{graphicx}
\usepackage{float}

\journal{Annals of Physics}
\begin{document}

\begin{frontmatter}

\title{Position Measurements in the de Broglie - Bohm Interpretation of
Quantum Mechanics}

\author{Gillie Naaman-Marom}
\author{Noam Erez}
\author{Lev Vaidman\corref{cor1}}
\ead{vaidman@post.tau.ac.il}
\cortext[cor1]{Tel. +972-3-6406014; Fax +972-3-6407932}
\address{Raymond and Beverly Sackler School of Physics and Astronomy\\
 Tel-Aviv University, Tel-Aviv 69978, Israel}

\begin{abstract}
The de Broglie - Bohm Interpretation of Quantum Mechanics assigns positions and
trajectories to particles.
We analyze the validity of a formula for the velocities
 of Bohmian  particles which makes the analysis of these
trajectories particularly simple. We apply it to four different types
of particle detectors and show that three types of the detectors lead
to ``surrealistic trajectories'', i.e., leave a trace where the Bohmian
particle was not present.
\end{abstract}


\end{frontmatter}

\section{ Introduction}

\label{intr}

It is remarkable, that a century after the discovery of quantum mechanics,
it seems that we are no closer to a consensus about its interpretation,
than we were in the beginning. The collapse of the quantum state at
the process of measurement which appears in all textbooks of quantum
theory does not have an unambiguous definition and a reasonable explanation.
The Einstein-Podolsky-Rosen (EPR) argument \cite{PhysRev.47.777}
that quantum mechanics is incomplete, together with Bell's proof \cite{bell1964einstein}
that it cannot be completed without some nonlocal phenomena made the
task of creating an understandable picture of quantum mechanics very
difficult. They showed that some radical changes in our classical
understanding of reality have to be made; e.g. constructing a physical
process of collapse \cite{Pearle76,PhysRevD.34.470}, accepting the
existence of parallel worlds \cite{vaidman2002many}, or adding non-local
hidden variables. Perhaps, the most attractive way to stay close to
the single-world classical picture of reality is to accept the de
Broglie-Bohm (dBB) interpretation \cite{Broglie28,Bohm52}
which includes  hidden variables, the positions of quantum particles moving on continuous trajectories
resembling classical physics. The dBB interpretation solves several
issues of the quantum measurement problem. It defines unambiguously
when a collapse of the quantum state takes place, it explains
in a very elegant way what happens in the EPR experiment and it restores
determinism in physics. Note, however, that it does not restore locality
in physics.

In this paper we will argue that a simple formula for Bohmian velocity
of a particle when it passes a region of the overlap of superposed
wavepackets can be widely applied. In particular, it helps analyzing
so-called ``surrealistic trajectories'' appearing in a setup proposed
by Englert \textit{et al.} \cite{englert1992surrealistic} which offer
a dramatic demonstration of the nonlocal nature of the dBB theory
and its consequences for the interpretation of measurements.

 The organization
of this paper is as follows. In section 2 we review
our approach to the dBB theory. In section 2.1 we present the ``velocity formula'',
in section 2.2 we apply it to the Stern-Gerlach (SG) experiment and in
section 2.3 to the EPR-Bohm experiment. In section 2.4 we analyze an
experiment in which an ``empty wave'' manifests itself by taking
the Bohmian particle from the other wave packet. In section 3 we
introduce four types of position measurements. In section 4 we present main results: our  analysis of
experiments which are devised  to monitor the trajectory of Bohmian position
 using the four types of position detectors.
In section 4.1 we consider particles with spin for which  the velocity formula is exact and shows unambiguously
the cases of surrealistic trajectories. In Section 4.2 we modify
the experiments, such that they allow the analysis of a spinless particle.
In this case the velocity formula is valid only approximately and
we compare it with an exact calculation of Bohmian position trajectories.
Since all our analysis was done with wave packets having unphysical
sharp edges, we perform, in Section 4.3, computer calculations for Gaussian
wave packets taking into account their time evolution. Comparing the
results of exact calculations and our toy model analysis demonstrates
effectiveness of our approach. The discussion of section 5 concludes
the paper.

\section{ The de Broglie-Bohm theory}

\subsection{  The formalism}

The dBB theory postulates that the quantum state
evolves according to the Schrodinger equation and never collapses. In addition, every particle (photons
currently not included) has definite position at all times and its
motion is governed in a simple way by the quantum state. The history
of the dBB is complex. De Broglie \cite{Broglie28} first presented it in 1927
and subsequently changed his position, presenting a significantly
different view \cite{deBroglie1956}. Bohm \cite{Bohm52} made a clear exposition
in 1952, although he never viewed
it as a proposal for a final theory, but rather as a way for developing
a new and better approach. The motions of ``particles'' in the
de Broglie and Bohm theories are identical, but the important difference
between these formulations is that de Broglie used an equation for
velocity (determined by the quantum state) as the guiding equation
for the particle, while Bohm used equation for acceleration (a la
Newton) introducing a ``quantum potential'' (likewise determined
by the quantum state). The version we find the most attractive was
advocated by Bell \cite{bell2004speakable} and it is closer to de
Broglie as the equation of motion is given in terms of the velocity.
An important aspect is that the only ``Bohmian'' variables in
this approach are particle positions (and variables like velocity that are defined in terms
 of positions), while all other variables, e.g.
spin, are described solely by the quantum state.

Originally, it was postulated that the initial distribution of Bohmian
positions of particles is according to probability law $|\psi|^{2}$
in configuration space. There are claims that this postulate is unnecessary
because in large systems (i.e., composed of many particles) even if
the Bohmian position in configuration space starts in a low probability
region, it will typically move to a high probability point very rapidly
\cite{PhysRev.89.458,Valentini2005}. Since in this approach we still
have to postulate something about initial Bohmian position -- it cannot
be where $\psi=0$ --we see little advantage of it relative to the
standard proposal, especially since the latter is applicable for all
systems.

We feel that dBB theory needs another postulate: \textit{our observations
supervene on Bohmian trajectories and not on a quantum state}. This
postulate is needed for defending dBB theory from so called ``idle
wheel attack'' according to which dBB theory is ``the Many-Worlds Interpretation in denial''
\cite{Valentini2010}. Indeed, the quantum state already includes
a structure corresponding to our reality. However, it also includes structures
corresponding to numerous other parallel realities and we need a
postulate to downgrade these structures. Then, a single structure
of trajectories will remain to manifest our single-world reality.
Myrvold \cite{Myrvold} has pointed out that the existence of this
postulate brings ``us'' or conscious observer into the definition
and thus apparently contradicts the claim of proponents of dBB that
it does not require the concept of an observer. However, one can argue
that the trajectories are the primary ontology and then it follows
that only they supervene on everything ``real'' and, in particular,
on our experience.

So, let us define the dBB theory we consider here. The ontology of
the theory consists of a quantum state, i.e., the wave function, and
the trajectories of all particles in three-dimensional space. Our
experiences supervene directly on the trajectories alone. The quantum
state evolves according to the Schrodinger equation (or, more precisely,
according to its relativistic generalization). It is completely deterministic,
the value of the wave function at any single time determines the wave
function at all times in the future and in the past. The evolution
of the particle positions is also deterministic. The velocity of each
particle at a particular time depends on the wave function and positions
of all the particles at that time. Namely, the velocity of particle
$i$ at time $t$ is given by:
 \begin{equation}
\dot{\boldsymbol{\textbf{r}}}_{i}(t)=
\textbf{Im}\frac{\hbar}{m}\left.\frac{\mathbf{\boldsymbol{\Psi}}^{\dagger}\left(\boldsymbol{r}_{1},\ldots,\boldsymbol{r}_{N},t\right)
\boldsymbol{\nabla}_{i}\boldsymbol{\Psi}\left(\boldsymbol{r}_{1},\ldots,\boldsymbol{r}_{N},t\right)}{\boldsymbol{\Psi}^{\dagger}
\left(\boldsymbol{r}_{1},\ldots,\boldsymbol{r}_{N},t\right)\boldsymbol{\Psi}\left(\boldsymbol{r}_{1},\ldots,\boldsymbol{r}_{N},t\right)}\right|_{\boldsymbol{r}_{i}=
\boldsymbol{\textbf{r}}_{i}(t)}.\label{eq:Bvel_preview}
\end{equation}
(We use Roman text font for Bohmian positions and bold font to signify
that $\mathbf{\boldsymbol{\Psi}}$ is a spinor.) An equivalent, but
suggestive statement of this law separates the procedure into two
steps \cite{Goldstein}. First, a \textit{conditional wave function}
of particle $i$ at particular time is defined by fixing the positions
of all other particle to be their Bohmian positions at that time:
\begin{equation}
\boldsymbol{\psi}_{i}(\boldsymbol{r}_{i},t)=
\boldsymbol{\Psi}\left(\boldsymbol{\textbf{r}}_{1},\ldots,\boldsymbol{r}_{i},\ldots,\boldsymbol{\textbf{r}}_{N},t\right).\label{psief}
\end{equation}
 Note that $\boldsymbol{\psi}_{i}$ still carries spin variables of
all the particles. The velocity of the $i$th particle is then:
\begin{equation}
\dot{\boldsymbol{\textbf{r}}}_{i}=
\frac{\hbar}{m}\textbf{Im}\left.\frac{\boldsymbol{\psi}_{i}^{\dagger}\boldsymbol{\nabla}\boldsymbol{\psi}_{i}}
{\boldsymbol{\psi}_{i}^{\dagger}\boldsymbol{\psi}_{i}}\right|_{\boldsymbol{r}_{i}=\boldsymbol{\textbf{r}}_{i}(t)}\label{rdot}.
\end{equation}
 This velocity formula ensures that a Bohmian particle inside a moving
wave packet which has group velocity $\bf{v}$ ``rides'' on the
wave with the same velocity, $\dot{\mathbf{r}}=\mathbf{v}$. For a
general wave, which does not consist of a single well- localized wavepacket,
as occurs, e.g., in a two slit experiment, the trajectory might be
complicated and we have to perform calculations (\ref{rdot}) to find
it.

An interesting case, appearing in numerous gedanken (as well as real)
experiments, is when the wave is a superposition of a few well localized
wave packets. Of course, one can still use formula (\ref{rdot})
and calculate the trajectories (with the help of a computer), but we are
looking for a simple method which gives at least the basic picture
without complicated calculations. One simple tool is the ``no-crossing
theorem'' which has been used extensively for so called ``surrealistic
trajectories'' examples \cite{englert1992surrealistic} but its
effectiveness is limited in practice to special classes of problems.
We argue that for finding Bohmian particle trajectory we can use,
instead, a simple formula which says that a velocity of a Bohmian
particle position located in the overlap of localized wave packets
is given by a weighted average of the velocities of these packets.
In many cases this formula provides exact results and in others it
yields a very good approximation of the Bohmian trajectory.

Consider the effective (conditional) wave wave function of a particle
consisting of a superposition of two localized wave packets $\psi=\psi_{{\rm A}}+\psi_{B}$
moving with well defined velocities $\mathbf{v}_{A}$ and $\mathbf{v}_{B}$
respectively and without significant spreading. Then, the velocity
formula for the Bohmian position of this particle is:
 \begin{equation}
\dot{\boldsymbol{\textbf{r}}}(t)=\frac{{\rho_{A}\mathbf{v}_{A}+\rho_{B}\mathbf{v}_{B}}}{{\rho_{A}+\rho_{B}}},\label{v}
\end{equation}
 where $\rho_{A}=\psi_{A}^{\dagger}(\boldsymbol{{\rm r}})\psi_{A}(\boldsymbol{{\rm r}})$,
$\rho_{B}(\boldsymbol{{\rm r}})=\psi_{B}^{\dagger}(\boldsymbol{{\rm r}})\psi_{B}(\boldsymbol{{\rm r}})$.

This formula is exact if the wave packets are entangled with orthogonal
spin states of this or another particle. Otherwise, it provides a good
approximation which sometimes improves when the wave packets are entangled
with other systems. If spreading of the wave packets is not negligible, then $\rho\mathbf{v}$ should be replaced by local currents in each wave packet.

\subsection{ Stern-Gerlach experiment}

One of the most striking quantum behaviors can be observed in spin
measurements which are performed using SG devices. The
most vivid example is the Greenberger-Horne-Zeilinger-Mermin setup \cite{greenberger1989going,mermin1990quantum,greenberger1990bell}
in which four alternative sets of measurements of spin $x$ and spin
$y$ components are considered on three entangled particles $A,B$
and $C$ in a Greenberger-Horne-Zeilinger state:
\begin{equation}
\boldsymbol{\Psi}_{\mathrm{GHZ}}=\frac{1}{\sqrt{2}}\left(|\uparrow_{z}\rangle|\uparrow_{z}\rangle|\uparrow_{z}\rangle-|\downarrow_{z}\rangle|\downarrow_{z}\rangle|\downarrow_{z}\rangle\right).\end{equation}
 The outcomes of the local measurements fulfill the following four
equations (each line corresponding to a different set of possible
measurements):
\begin{eqnarray}
\sigma_{1x}\sigma_{2x}\sigma_{3x} & = & -1,\nonumber \\
\sigma_{1x}\sigma_{2y}\sigma_{3y} & = & +1,\\
\sigma_{1y}\sigma_{2x}\sigma_{3y} & = & +1,\nonumber \\
\sigma_{1y}\sigma_{2y}\sigma_{3x} & = & +1.\nonumber
\end{eqnarray}
 Considering the product of these equation we immediately see that
there is no solution for outcomes of local measurement which fulfills
all of them. Thus, Einstein's hope that quantum mechanics can be completed
such that all ``elements of reality'' will have definite values
cannot be accomplished. In the dBB picture, indeed, spin components
have no definite values, nevertheless, everything is deterministic,
so the outcome of any SG experiment is fixed prior to experiment.
Let us see how it happens.

In our simplified model, the initial quantum state of the particle
is an equal weight wave in the form of a thin (in $x$ direction)
box with negligible spreading moving in the $x$ direction, with a
pure spin state. It has no spatial dependence in the $y$ direction,
so we will not mention it. We arrange that the wave packet reaches
the origin at time $t=0$:
\begin{equation}
\psi(t)=e^{ik(x-\frac{v}{2}t)}\chi(x-vt,z)(\alpha|\uparrow\rangle+\beta|\downarrow\rangle),\label{psiin}
\end{equation}
 where
 \begin{equation}
\chi(x,z)\equiv\begin{cases}
\frac{1}{{\sqrt{\epsilon a}}}~, & |x|<\frac{\epsilon}{2},~|z|<\frac{a}{2},\\
0, & {\rm otherwise}.\end{cases}\label{chi}
\end{equation}
 We will consider below also an exact solution for a Gaussian wave
function in order to show that our toy model faithfully represents
the behavior of the system. In the simple model, we can immediately
see that the Bohmian particle rides the wave, i.e., the position of
any Bohmian particle placed in the support of the particle's wave
function is described, for $t<0$, by:
\begin{equation}
\mathbf{r}(t)=(\mathrm{x}_{0}+vt,\mathrm{z}_{0}).
\end{equation}

In order to avoid subtleties of a real SG experiment \cite{Home},
we consider a simplified gedanken model of it. At time $t=0$, when
the wave packet reaches the origin, a spin-dependent momentum kick
in the $z$ direction is provided without changes in other directions.
This results in a vertical component of the wave packet velocity of
magnitude $\pm u$ for the respective spin components. The wave function at $t>0$ is described by
\begin{equation}
e^{ik_{x}(x-\frac{v}{2}t)}\left(\alpha e^{ik_{\uparrow z}(z-\frac{u}{2}t)}\chi(x-vt,z-ut)|\uparrow\rangle+\beta e^{ik_{\downarrow z}(z+\frac{u}{2}t)}\chi(x-vt,z+ut)|\downarrow\rangle\right).\label{psiout}
\end{equation}
 In order to keep the equations tractable, we will omit from now on
everything related to the $x$ axis. Then, the wave function (\ref{psiout})
will have the following form:
 \begin{equation}
\alpha e^{ik_{z}(z-\frac{u}{2}t)}\chi(z-ut)|\uparrow\rangle+\beta e^{-ik_{z}(z+\frac{u}{2}t)}\chi(z+ut)|\downarrow\rangle,\label{psioutnox}
\end{equation}
 where $k_{z}=k_{\uparrow z}=-k_{\downarrow z}$ and
 \begin{equation}
\chi(z)\equiv\begin{cases}
\frac{1}{{\sqrt{a}}}~, & ~|z|<\frac{a}{2},\\
0, & {\rm otherwise}.\end{cases}\label{chi1}
\end{equation}
 It is easy to calculate the motion of Bohmian particle after the
momentum kick. During the time the particle is located in the overlap
of the two wave packets, the $z$ component of the velocity is
\begin{equation}
\mathrm{v}_{z}=\dot{\mathrm{z}}=\frac{{\rho_{\uparrow}~u+\rho_{\downarrow}~(-u)}}{{\rho_{\uparrow}+\rho_{\downarrow}}}=\left(|\alpha|^{2}-|\beta|^{2}\right)u.\label{vz}
\end{equation}
 At some time, the particle will leave the (changing) overlap area
and from that moment will ride one of the wave packets. Thus, to find
the result of the measurement, we have to calculate what will happen
first: will the particle be ``overtaken'' by the bottom of the
$\uparrow$ wave packet (and from then on will ride the $\downarrow$ wave packet)
or will it be overtaken by the top of the $\downarrow$ wave packet (and will
subsequently ride the $\uparrow$ wave packet)? To this end we assume that
the particle moves with velocity (\ref{vz}) and calculate the hypothetical
times of these events. The earlier one  is the one that
will actually take place. These times can be found from the following
equations:
\begin{eqnarray}
\mathrm{z}_{0}+\left(|\alpha|^{2}-|\beta|^{2}\right)ut & = & -\frac{a}{2}+ut,\label{overlap1}\\
\mathrm{z}_{0}+\left(|\alpha|^{2}-|\beta|^{2}\right)ut & = & \frac{a}{2}-ut.
\end{eqnarray}
 The equality of times corresponds to
 \begin{equation}
\mathrm{z}_{0}=\left(\frac{1}{2}-|\alpha|^{2}\right)a.
\end{equation}
 Thus, any Bohmian particle in the upper $|\alpha|^{2}$ fraction of
the wave will be ``caught'' by the $\uparrow$ wave packet, so that the
particle will be detected in the upper beam which corresponds to the
outcome of the experiment $\sigma_{z}=1$. Similarly, any particle
in the lower $|\beta|^{2}$ part of the wave will end up in the
$\downarrow$ wave packet and the outcome of the experiment will be $\sigma_{z}=-1$.
In Fig.1 a particular example with $\alpha=\sqrt{\frac{2}{5}},\beta=\sqrt{\frac{3}{5}},~z_{0}=\frac{a}{5}$
is presented. From Eq.(\ref{overlap1}) we see that the Bohmian position
will be caught by wave packet $\uparrow$ at $t=\frac{{3a}}{{8u}}$. It will
ride at the point located $\frac{a}{4}$ below the center of the wave
packet $\uparrow$.

\begin{figure}[t]
  \includegraphics[width=6cm]{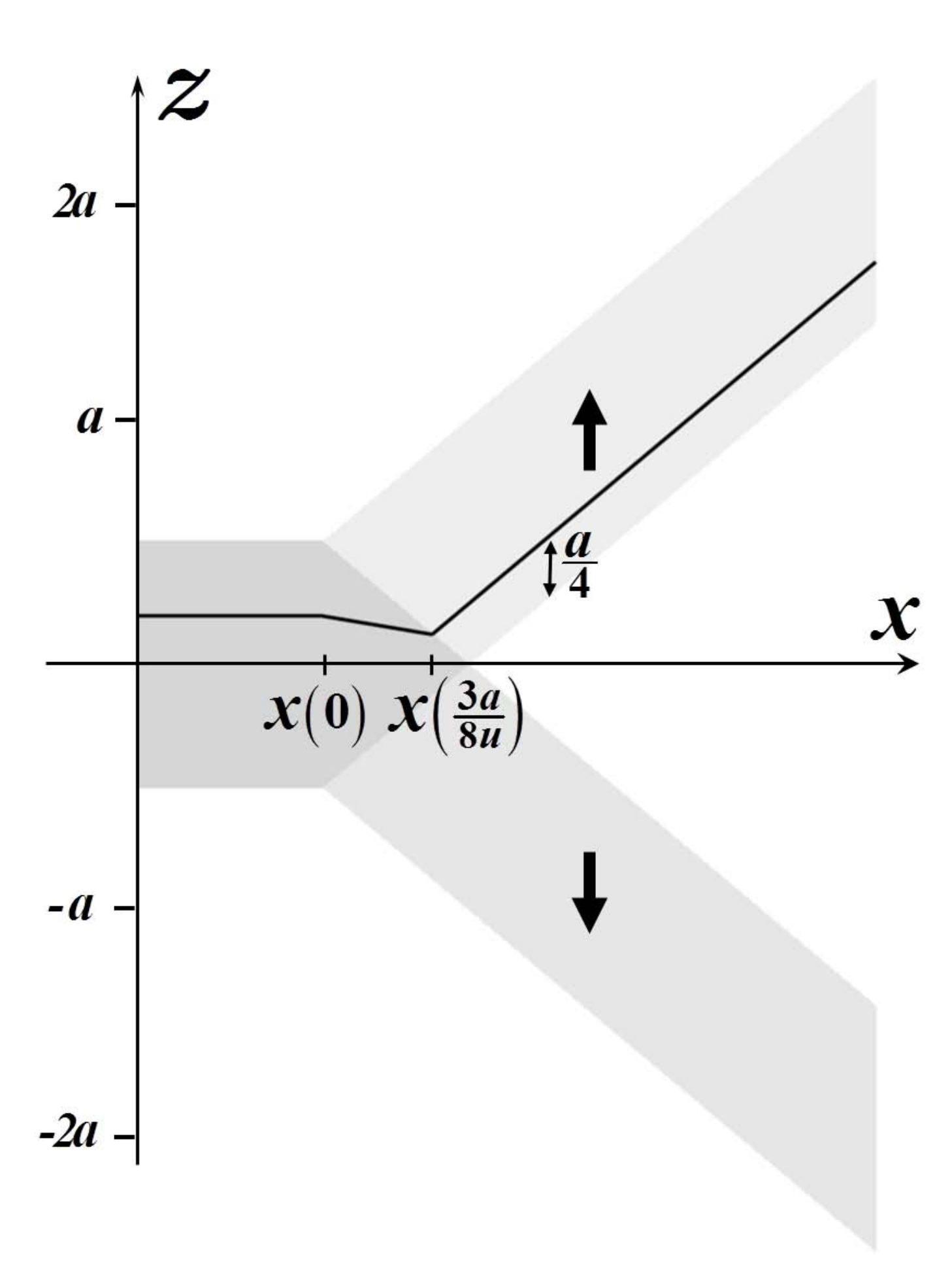}\\
  \vspace {-5pt}
    \caption{Bohmian trajectory in the SG device for a quantum particle in a spin state \break $\sqrt{\frac{2}{5}}|\uparrow\rangle+\sqrt{\frac{3}{5}}|\downarrow\rangle$ and initial Bohmian position $z_{0}=\frac{a}{5}$. Darker shade of grey indicates higher amplitude of the wave function $|\psi|$ at  times  when ${\rm x}(t)=x$, i.e. $|\psi(x,z,\frac{x-{\rm x}_0}{v})|$ .}
\end{figure}

In spite of the fact that the quantum wave has components corresponding
to two outcomes, the dBB theory shows in a deterministic and very
simple way how a single outcome emerges at each experimental run.
The quantum state (\ref{psiin}) and the initial Bohmian position
$z_{0}$ determine the outcome of the experiment.

In fact this is not the whole story: $\boldsymbol{\psi}(0)$ and $\mathbf{r}(0)$
alone are not enough to decide the outcome of the spin component measurement:
$\sigma_{z}=1$ or $\sigma_{z}=-1$. We have to specify the design
of our measuring device. To show this in the most vivid way,  consider the case of $|\alpha|=|\beta|$. We can, without changing anything about the
initial quantum state and Bohmian position of our particle, change the SG device such that it will provide the  spin-depended momentum kick
in the opposite direction. (It is instructive that a naive rotation by $\pi$ of the SG device does not cause the desired change \cite{Ghirardi}.) The wave function of the particle for $t>0$
will then be
 \begin{equation}
\alpha e^{-ik(z+\frac{u}{2}t)}\chi(z+ut)|\uparrow\rangle+\beta e^{ik(z-\frac{u}{2}t)}\chi(z-ut)|\downarrow\rangle.\label{psioutnox+}
\end{equation}
 The Bohmian trajectory will not be altered due to this change,
but this time, detection of the particle in the upper beam will correspond
to the outcome $\sigma_{z}=-1$.

\subsection{ The EPR experiment}

The ``contextuality'' of the measurement in the dBB theory discussed
above does not help to resolve the EPR paradox. Let us consider the
Bohm version of the EPR experiment with the SG devices
described above. Two particles in distant
laboratories, 1 and 2, are  initially in a maximally entangled state of their
spin variables and  in the product spatial state. We denote positions of the particles relative to their local frames
of reference:
 \begin{equation}
\frac{1}{\sqrt{2}}\chi(z_{1})\chi(z_{2})(|\uparrow\rangle_{1}|\downarrow\rangle_{2}-|\downarrow\rangle_{1}|\uparrow\rangle_{2}).
\end{equation}
 If we perform a measurement on particle 1 using SG device
described above, the quantum state will become:
 \begin{equation}
\frac{1}{\sqrt{2}}\chi(z_{2})\left(e^{ik(z_{1}-\frac{u}{2}t)}\chi(z_{1}-ut)|\uparrow\rangle_{1}|\downarrow\rangle_{2}-e^{-ik(z_{1}
+\frac{u}{2}t)}\chi(z_{1}+ut)|\downarrow\rangle_{1}|\uparrow\rangle_{2}\right).
\end{equation}
A simple straightforward analysis, similar to the case of a single
spin$-\frac{1}{2}$ particle, shows that if the Bohmian position of
particle 1 has initially ${\rm z}_{1}>0$, the particle will move
up. Indeed, the measurement interaction splits the wave packet into
a superposition of two wave packets of equal weight. While the Bohmian
position is inside the overlap of the two packets, it has zero component
of velocity in the $z$ direction and it is the top of the wave packet  $\downarrow$
which will be the first to reach it. There is a symmetry between particle
1 and particle 2, so if initially ${\rm z}_{2}>0$ and we perform
SG measurement on particle 2 instead, it will also move
up. Nothing prevents us from initially having both ${\rm z}_{1}>0$
and ${\rm z}_{2}>0$, but they cannot both move ``up'' since in
the EPR-Bohm experiment the outcomes have to be anticorrelated.

To simplify the analysis, let us assume that we first perform a measurement
on particle 1 and then on particle 2. Initially, and until the Bohmian
position of the first particle leaves the overlap of its two wave
packets, $\chi({\text{z}}_{1}-ut)=\chi({\text{z}}_{1}+ut)$, so the
conditional wave function of the particle 2 is:
\begin{equation}
\frac{1}{\sqrt{2}}\chi(z_{2})\left(e^{ik({\text{z}}_{1}-\frac{u}{2}t)}|\uparrow\rangle_{1}|\downarrow\rangle_{2}-e^{-ik_{z}({\text{z}}_{1}
+\frac{u}{2}t)}|\downarrow\rangle_{1}|\uparrow\rangle_{2}\right).
\end{equation}
 But when the Bohmian position of particle 1 leaves the overlap area,
$\chi({\text{z}}_{1}+ut)=0$, the conditional wave function of the
second particle collapses to
\begin{equation}
\chi(z_{2}) e^{ik({\text{z}}_{1}-\frac{u}{2}t)}|\uparrow\rangle_{1}|\downarrow\rangle_{2}.
\end{equation}
 For such a wave function, in the SG measurement of particle 2 all its Bohmian positions  will
move down and thus, the spin measurements will indeed show anticorrelation.

What resolves the EPR paradox in the dBB picture is its nonlocality.
There is a real action at a distance. By performing or not performing
a measurement on particle 1, we can change the outcome the SG
measurement on particle 2 which takes place in a space-like separated
region. The collapse of the conditional wave function of
particle 2 happens instantaneously at the moment the Bohmian position
of particle 1 leaves the overlap area of its two wave packets.

\subsection{ Action of an Empty wave}

The picture presented above, apart from its non-locality, is very
natural and attractive. A less intuitive feature of Bohmian trajectories
shows up when we bring together two wave packets, one of which containing the
Bohmian position. Let us consider again the
wave function of the particle after leaving the SG device
(\ref{psiout}) and arrange at time $T$ a spin dependent momentum
kick in the opposite direction twice as strong as the previous
one. At time $t=2T$, the two wave packets will fully overlap. In
order to keep the notation compact, we will define this time as $t=0$
(i.e. set the clocks back by $2T$). Then, for $t>-T$, the wave function
will be
\begin{equation}
\alpha e^{-ik(z+\frac{u}{2}t)}\chi(z+ut)|{\uparrow}\rangle+\beta e^{ik(z-\frac{u}{2}t)}\chi(z-ut)|{\downarrow}\rangle.\label{psioutnox1}
\end{equation}

 Consider again the example  in which Bohmian trajectory started
at $\mathrm{z}_{0}=\frac{a}{5}$ using the velocity formula (\ref{v}). Bohmian
particle will ride the wave packet $\uparrow$ at the location $\frac{a}{4}$
below the center of the wave packet, see Fig. 2. The two wave packets fully overlap
at time $t=0$. Shortly before this, at time $t=-\frac{3a}{8u}$,
when $\mathrm{z}=\frac{a}{8}$, the Bohmian trajectory enters the overlap
of the two wave packets. For the region of the overlap, formula (\ref{v})
yields $\dot{\mathrm{z}}=(|\beta|^{2}-|\alpha|^{2})u$. Thus, at time
$t=\frac{a}{4u}$, when $\mathrm{z}=\frac{a}{4}$, the particle leaves the
overlap region and rides the center of the wave packet $\downarrow$, see Fig. 2.
\vspace {-12pt}
\begin{figure}[H]
  \includegraphics[width=9cm]{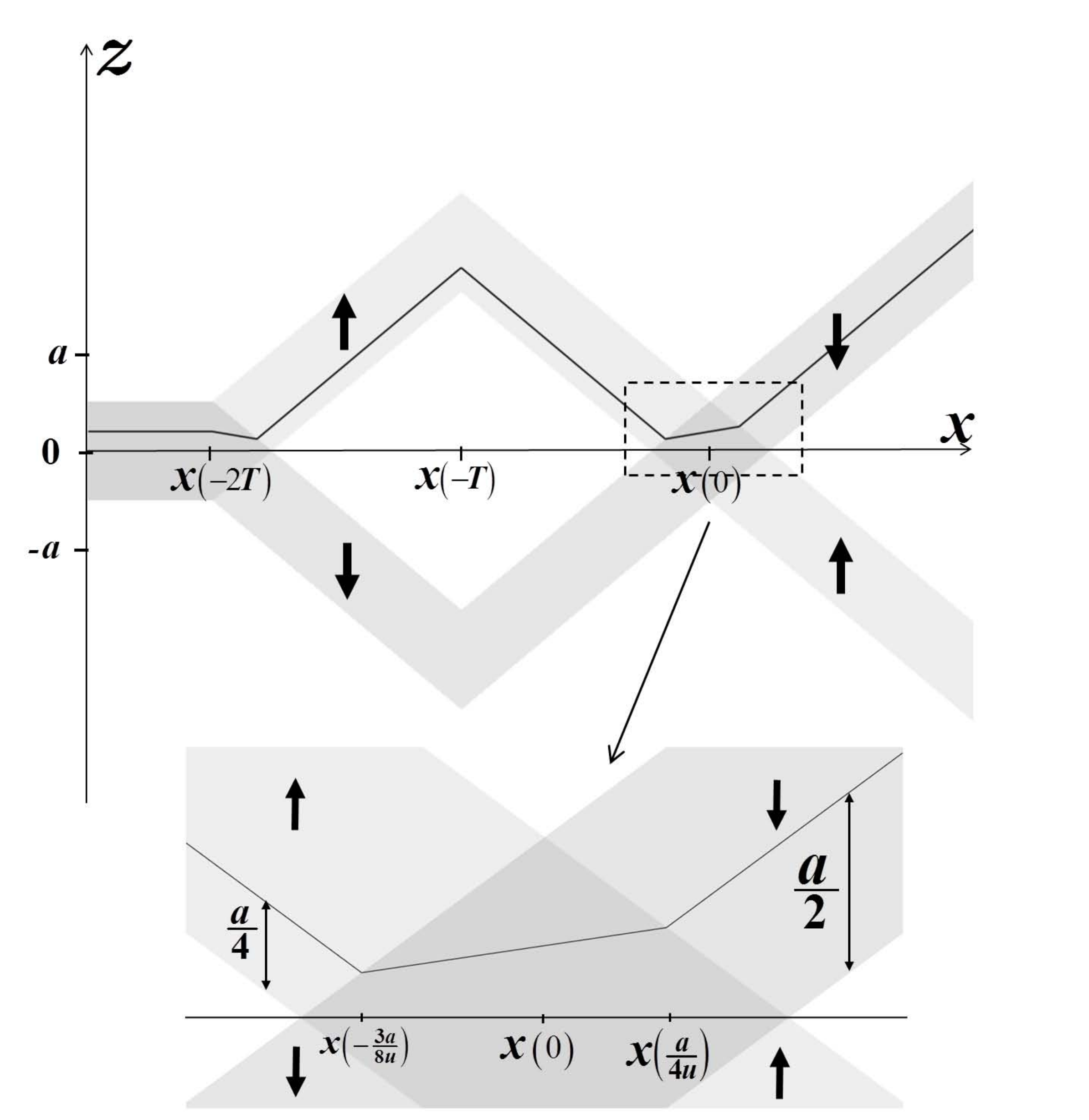}\\
  \vspace {-18pt}
       \caption{ The Bohmian trajectory of a particle $\rm z(x)$ and the amplitude of the wave function $|\psi(x,z,\frac{x-{\rm x}_0}{v})|$ (the shade of the grey color). The Bohmian particle changes the direction of its motion in the area of overlapping wave packets where electromagnetic field is not present.}
\end{figure}

In the process of switching from one wave packet to another, the Bohmian
position accelerates in the region where there are no physical fields.
For de Broglie and Bell this is an example of an action of
a ``pilot wave'' while for Bohm this is an interaction due to
a quantum potential. This phenomenon persists even if we
try to monitor the trajectory using apparently legitimate position
detectors. Since the detectors show one trajectory, whereas
the Bohmian particle actually takes, according to the theory, a different
trajectory, the latter were named ``surrealistic trajectories''
\cite{englert1992surrealistic}.

We will analyze this phenomenon considering
four different types of position detectors. In all cases, the detector
records the presence of a particle in its vicinity  via certain change
in a quantum state of a detector particle which will later be amplified
to a macroscopic recording in the detector. The differences between
the detectors are in the type of microscopic record of the detector
particle. Note, that  the amplification to a macroscopic record performed too early  might change the action of  a detector, so it has to be done after relevant evolution of the particle's wave.

\section{ Position detectors}

To avoid cumbersome notation and confusion between the measured particle
and detector particle we will call, from now on, the detector particle
a ``neutron''. This is just a name, we do not say that  neutrons are typically used in position detectors .

\begin{figure}[H]
  \includegraphics[width=4.5cm]{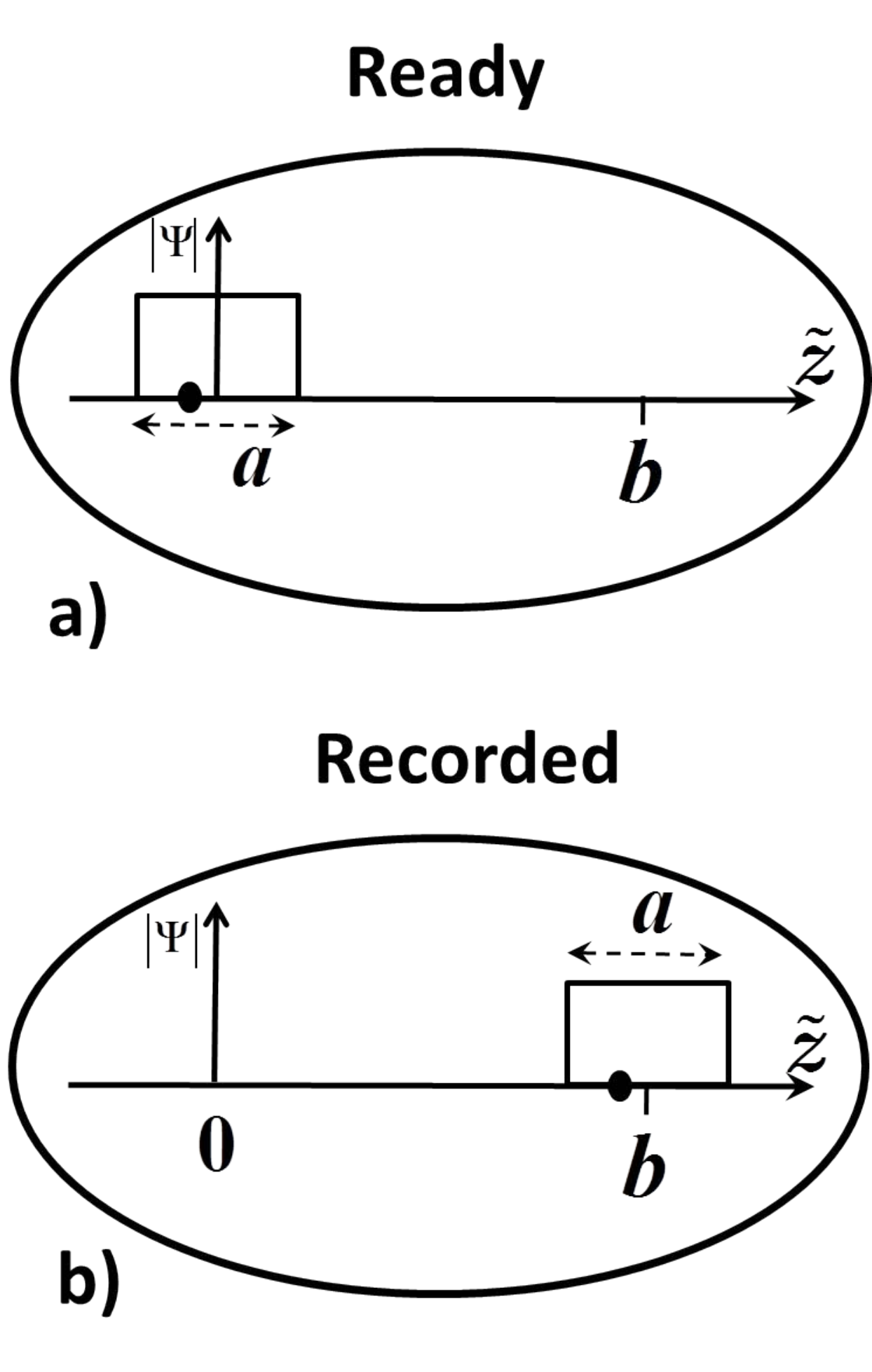}\\ \vspace{-6pt}
    \caption{ Bohmian position detector. The wave function  and the Bohmian position (black dot) of the neutron. (a)    Before the measurement as well as after the measurement, in case the particle was not present. (b) The case when the particle is detected. The Bohmian position specifies the result of the measurement.}
\end{figure}

\subsection{ Type {\rm (i)}: Bohmian position detector}

 This is a model of a standard
measuring device which shows its outcome by a pointer, see Fig. 3. The neutron
is initially at rest in a well localized wave packet. We will consider  its wave function in one dimension, $\chi(\tilde{z})$, and model it  as above (\ref{chi1}). We will use ``$~\tilde{~}~$'' to
signify variables of the measuring device. The interaction is such
that if our measured particle is present at the detector, the neutron
is quickly shifted by a distance $b$ larger than the width of its
wave packet:
\begin{equation}
\chi(\tilde{z})\rightarrow \chi(\tilde{z}-b).\label{psiout2}
\end{equation}
 We call it a ``Bohmian position detector'' because the information
about the result of the measurement is written (apart from the quantum
state of the neutron) in the Bohmian position of the neutron: ``particle
is not present'' corresponds to the Bohmian position ${\rm \tilde{z}}\in[-\frac{a}{2},\frac{a}{2}]$
and ``particle present'' corresponds to the Bohmian position ${\rm \tilde{z}}\in[b-\frac{a}{2},b+\frac{a}{2}]$.

\begin{figure}[h]
  \includegraphics[width=4.5cm]{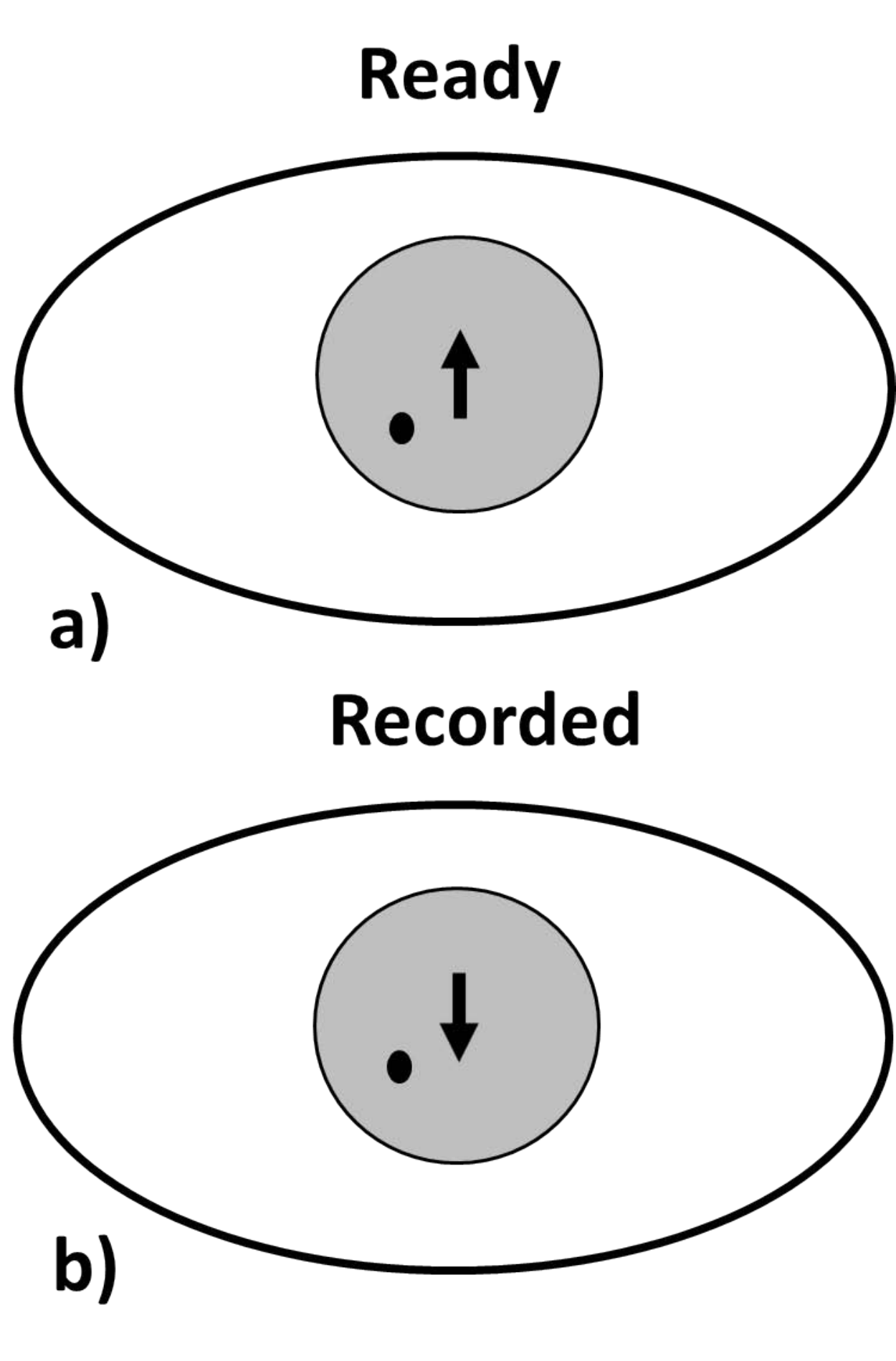}\\ \vspace{-6pt}
    \caption{ Spin detector.  The wave function and the Bohmian position of the neutron. (a)  Before the measurement as well as after the measurement, in case the particle was not present. (b) The case when the particle is detected.  The Bohmian position of the neutron is not changed.}
\end{figure}

\subsection{  Type {\rm (ii)}: spin detector}

 The spin of the neutron flips if the particle
is present, see Fig. 4:
 \begin{equation}
|\tilde{\uparrow}\rangle\rightarrow|\tilde{\downarrow}\rangle.\label{spindet}
\end{equation}
 Note that in this case, the Bohmian position of the neutron does not ``see''
the particle. It does not change its trajectory due to the presence
or absence of the particle. Only later, when the spin of the neutron
is observed, will the Bohmian trajectory of the neutron change
accordingly. But we assume that this stage happens long after the
phenomenon we discuss takes place.

\begin{figure}[h]
  \includegraphics[width=4.5cm]{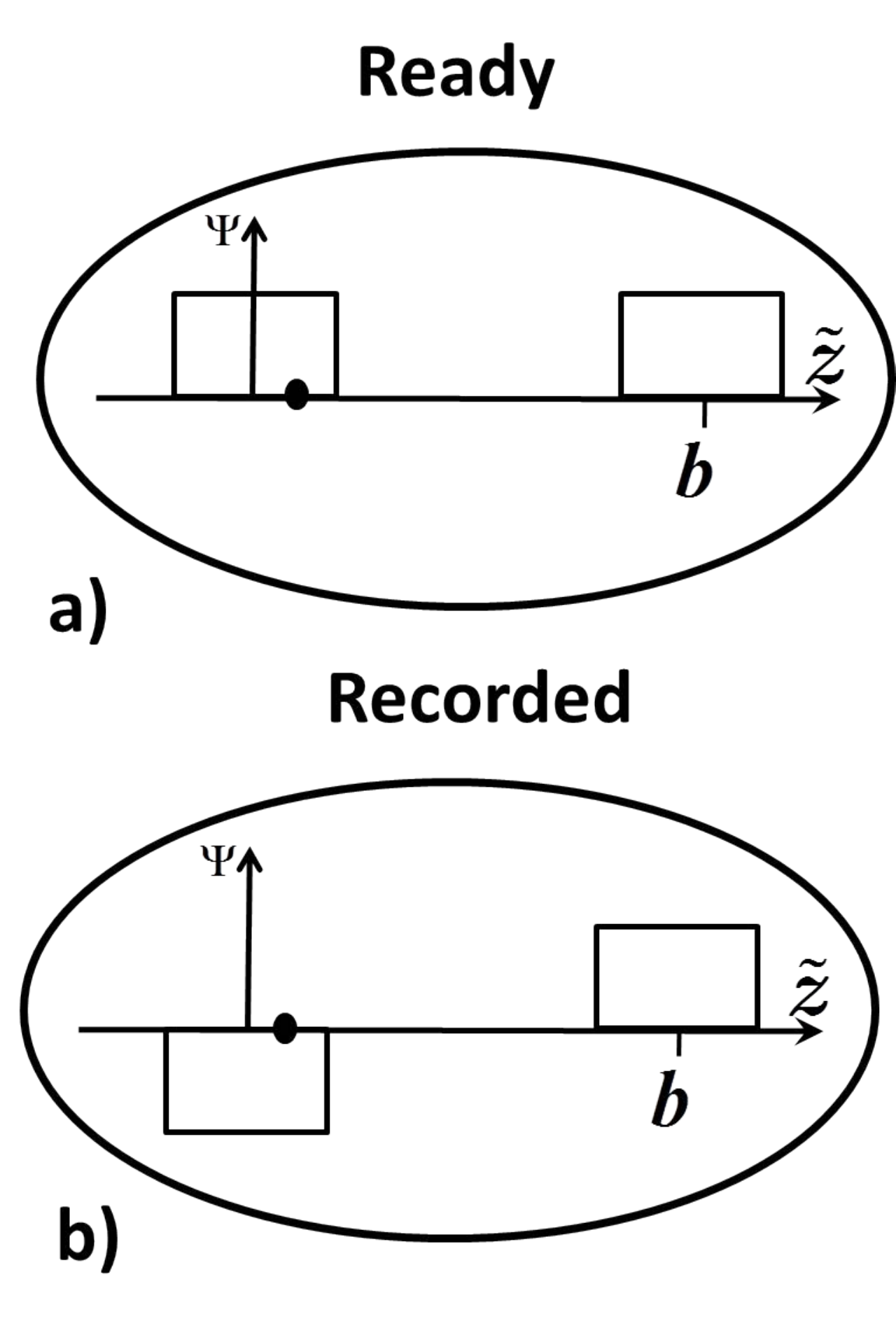}\\  \vspace{-6pt}
    \caption{ Phase detector.   The wave function and the Bohmian position of the neutron. (a)  Before the measurement as well as after the measurement, in case the particle was not present. (b) The case when the particle is detected.  The Bohmian position of the neutron is not changed.}
\end{figure}

\subsection{ Type {\rm (iii)}: Phase detector}

The neutron changes the relative
phase between the spatially separated parts of its wave function depending
on the presence of the observed particle see Fig. 5. In this detector, the neutron
is in a superposition of two well localized wave packets: one, $\chi(\tilde{z})$,
is near the place where the measured particle might be and the other,
$\chi(\tilde{z}-b)$, is  far away. Initially it is in
a superposition $\frac{1}{\sqrt{2}} \Big(\chi(\tilde{z})+\chi(\tilde{z}-b) \Big)$.
The measurement interaction is such that if the particle is present,
the potential near it leads to a relative phase of $\pi$ and the
neutron state changes as follows:
 \begin{equation}
\frac{1}{\sqrt{2}}\Big(\chi(\tilde{z})+\chi(\tilde{z}-b)\Big ) \rightarrow \frac{1}{\sqrt{2}}\Big(-\chi(\tilde{z})+\chi(\tilde{z}-b)\Big ).\label{phasedet}\end{equation}
 Otherwise, it remains unchanged. Again, the Bohmian position of the
neutron does not ``see'' the particle. It does not change its
trajectory due to the presence or absence of the particle. Only later,
when we bring together the two wave packets of the neutron together
for measuring the phase, will the Bohmian trajectories of the neutron
change accordingly.

\begin{figure}[H]
  \includegraphics[width=4.5cm]{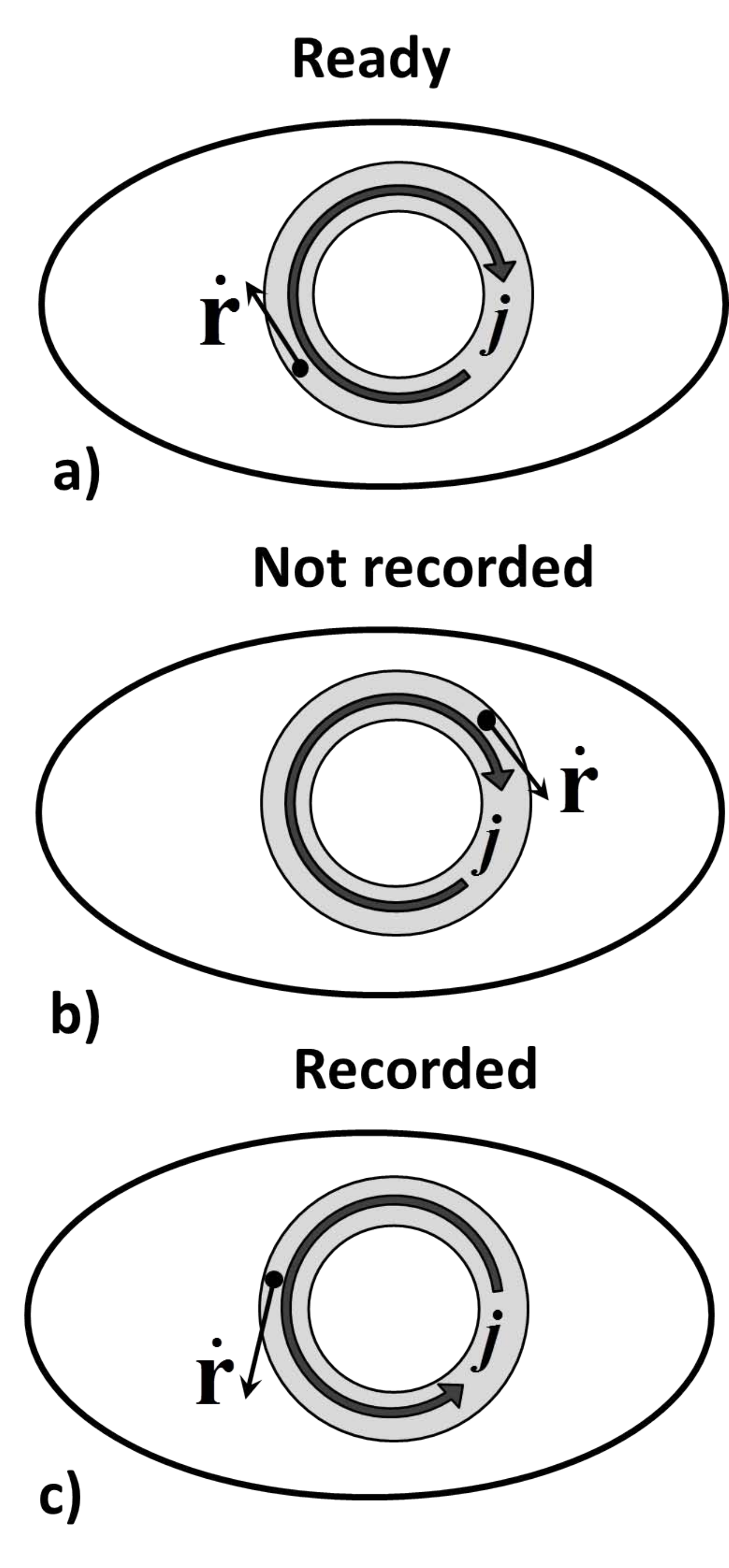}\\ \vspace{-6pt}
    \caption{ Bohmian velocity detector. (a) The current of the neutron is initially in the clockwise direction. (b) The current remains the same if the particle is not found. (c) The current changes its direction if the particle is present.  The Bohmian particle  velocity  has the same direction as the current at its location. The result of the measurement is ``recorded'' in the direction of Bohmian particle velocity (as well as in the direction of its current), but not in its Bohmian position, since the latter is random whatever the result is.}
\end{figure}

\subsection{  Type {\rm (iv)}: Bohmian velocity detector}

 The quantum state of the neutron
is spread out uniformly on a ring of radius $R$ located near the
place where presence of our particle is tested, see Fig. 6. The  wave
function of the neutron is effectively one-dimensional. Initially, the wave function of the neutron and its Bohmian position rotate
in clockwise direction and this motion remains unchanged if the particle
is not present. If it is present, wave function  of the neutron changes:
\begin{equation}
\frac{1}{\sqrt{2\pi R}}e^{-i\tilde{k}R(\tilde{\theta}+\frac{\omega}{2}t)}\rightarrow\frac{1}{\sqrt{2\pi R}}e^{i\tilde{k}R(\tilde{\theta}-\frac{\omega}{2}t)}.\label{BohmVdet}
\end{equation}
 Thus, the current and the Bohmian velocity change the direction of rotation. The outcome of the measurement is recorded in the motion of
the Bohmian position of the neutron: it moves with velocity $\omega R$
in the clockwise direction if there is no particle and it moves with
the same velocity in the counterclockwise direction if the particle is present.
This detector, however, is not a Bohmian position detector: the Bohmian position of the neutron
 can be anywhere on the ring whether the particle is present or not.

\section{ Results}

\subsection{ Particle with spin.}

We will now try to observe the action of an empty wave using position
detectors. We will start the analysis with Bohmian detector (i). We
put our detector in the lower arm of the device, Fig. 7a. The wave
function of the particle and the neutron for $t>-T$ is
\begin{equation}
\alpha e^{-ik(z+\frac{u}{2}t)}\chi(z+ut)\chi(\tilde{z})|\uparrow\rangle+\beta e^{ik(z-\frac{u}{2}t)}\chi(z-ut)\chi(\tilde{z}-b)|\downarrow\rangle.\label{psioutnox2}
\end{equation}
 Consider again the initial Bohmian position of the particle ${\rm z_{0}}=\frac{a}{5}$.
Since at time $t=-T$ the Bohmian position of the particle is inside
the  wave packet in the upper arm of the interferomenter, the conditional wave function of the neutron
remains $\chi(\tilde{z})$ and thus, its Bohmian position remains
unchanged: ${\rm \tilde{z}}\in[-\frac{a}{2},\frac{a}{2}]$. Then,
at time $t>-T$, the conditional wave function of the particle is:
\begin{equation}
e^{-ik(z+\frac{u}{2}t)}\chi(z+ut)|\uparrow\rangle.\label{psiout55}
\end{equation}
 Therefore, the Bohmian position of the particle will continue to
ride wave packet moving towards detector $A$ at the location $\frac{a}{4}$ below its center. This corresponds to our expectation: the particle
detected in $A$ does not leave a record at the detector located on
another path: we observe no surrealistic trajectories.

Now consider spin detector (ii). The neutron starts in the state $|\tilde{\uparrow}\rangle$ and
flips its spin at time $t=-T$ if the particle passes through the lower arm. Then, the wave function of the particle and the neutron for
$t>-T$ is:
 \begin{equation}
\alpha e^{-ik(z+\frac{u}{2}t)}\chi(z+ut)|\tilde{\uparrow}\rangle)|\uparrow\rangle+\beta e^{ik(z-\frac{u}{2}t)}\chi(z-ut)|\tilde{\downarrow}\rangle)|\downarrow\rangle.\label{psioutnox33}
\end{equation}
 The presence of the spin detector changes nothing regarding the motion
of the Bohmian particle relative to the case without the detector (\ref{psioutnox1}):
before time $t=-T$ the wave packets $A$ and $B$ were entangled
to the spin of the particle and now we add entanglement to the spin
of the neutron. The spatial density of the particle remains the same
and formula (\ref{v}) holds exactly. The Bohmian particle does change
its direction at the area of the overlap and, as in
the case without detector, it moves up towards
detector $B$, see Fig. 7b.  Here we can see the meaning of the requirement that the amplification to a macroscopical record should take place ``later''. The validity of Eq. (\ref{psioutnox33}) relies on the fact that the amplification did not take place. The amplification had to be done after the   the particle wave packets overlap,  $t>0$. At the moment the Bohmian position of the particle leaves the overlap area, 
the neutron conditional wave function collapses to $|\tilde{\downarrow}\rangle$. This naively corresponds to the neutron recording the particle in 
the lower arm in spite of the fact that the Bohmian trajectory passed through the
upper arm.

\begin{figure}[H]
   \vspace{-20pt}
   \includegraphics[width=6.3cm]{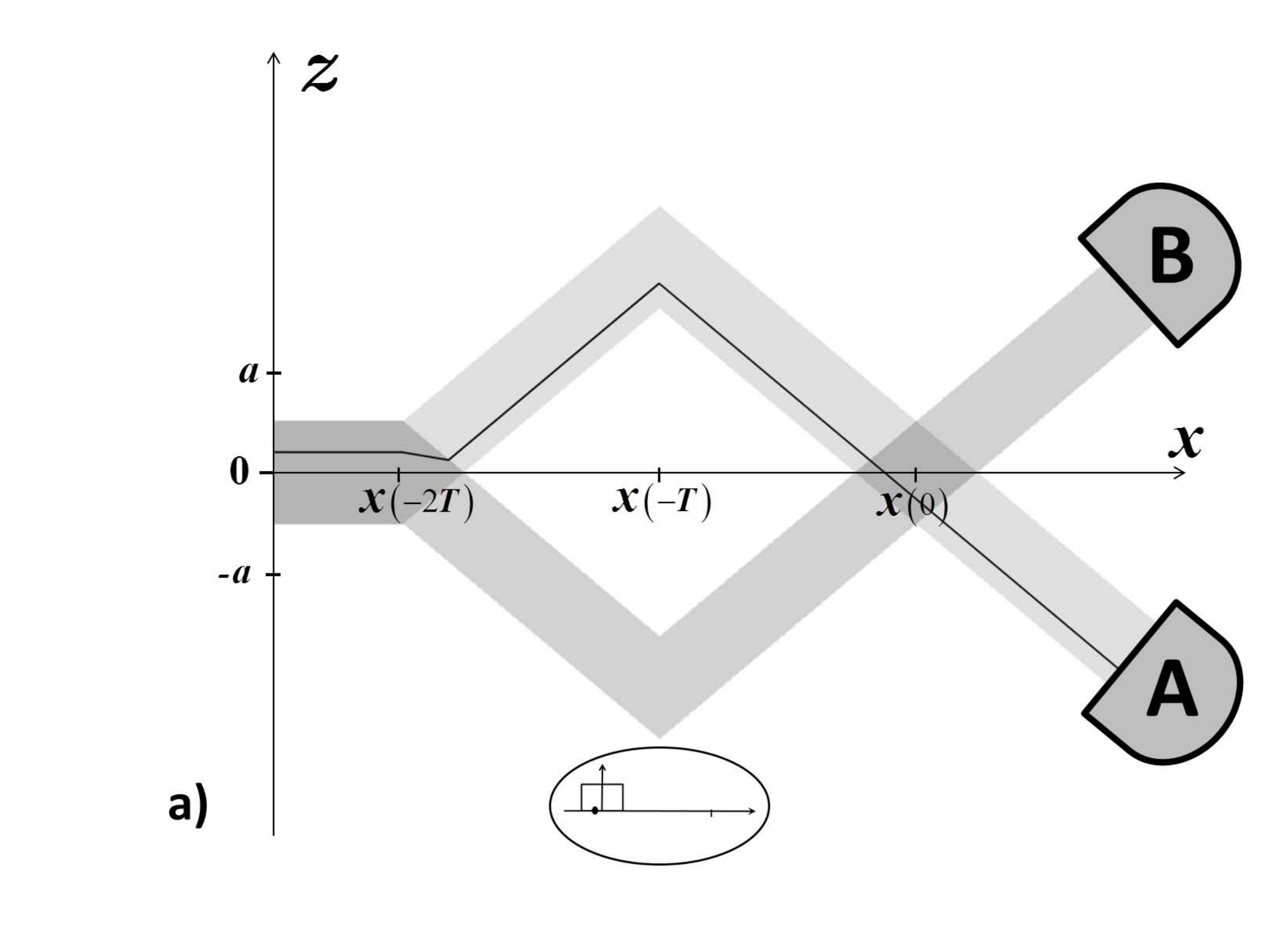}\\
  \vspace{-10pt}
   \includegraphics[width=6.3cm]{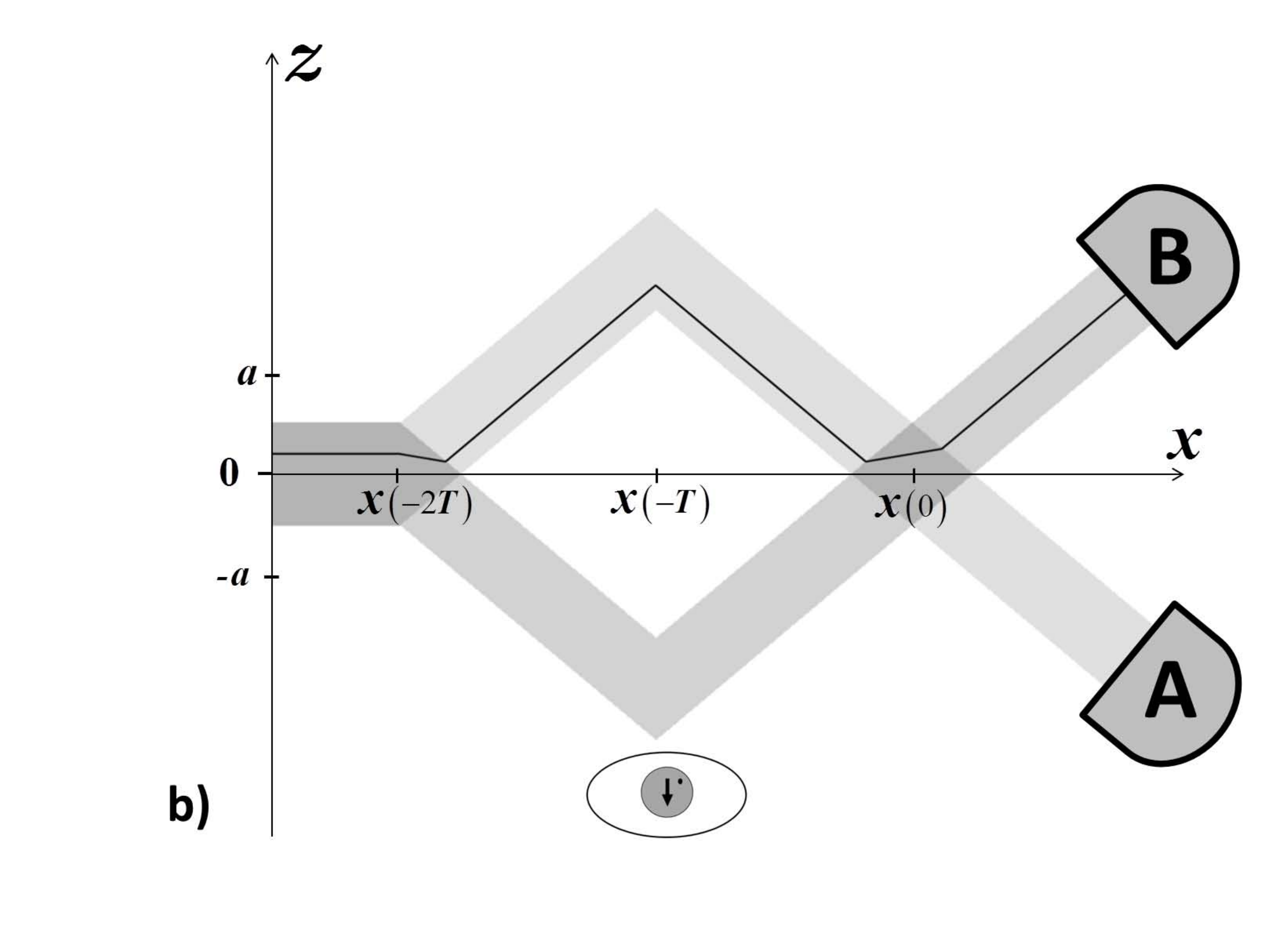}\\ \vspace{-8pt}
     \includegraphics[width=6.3cm]{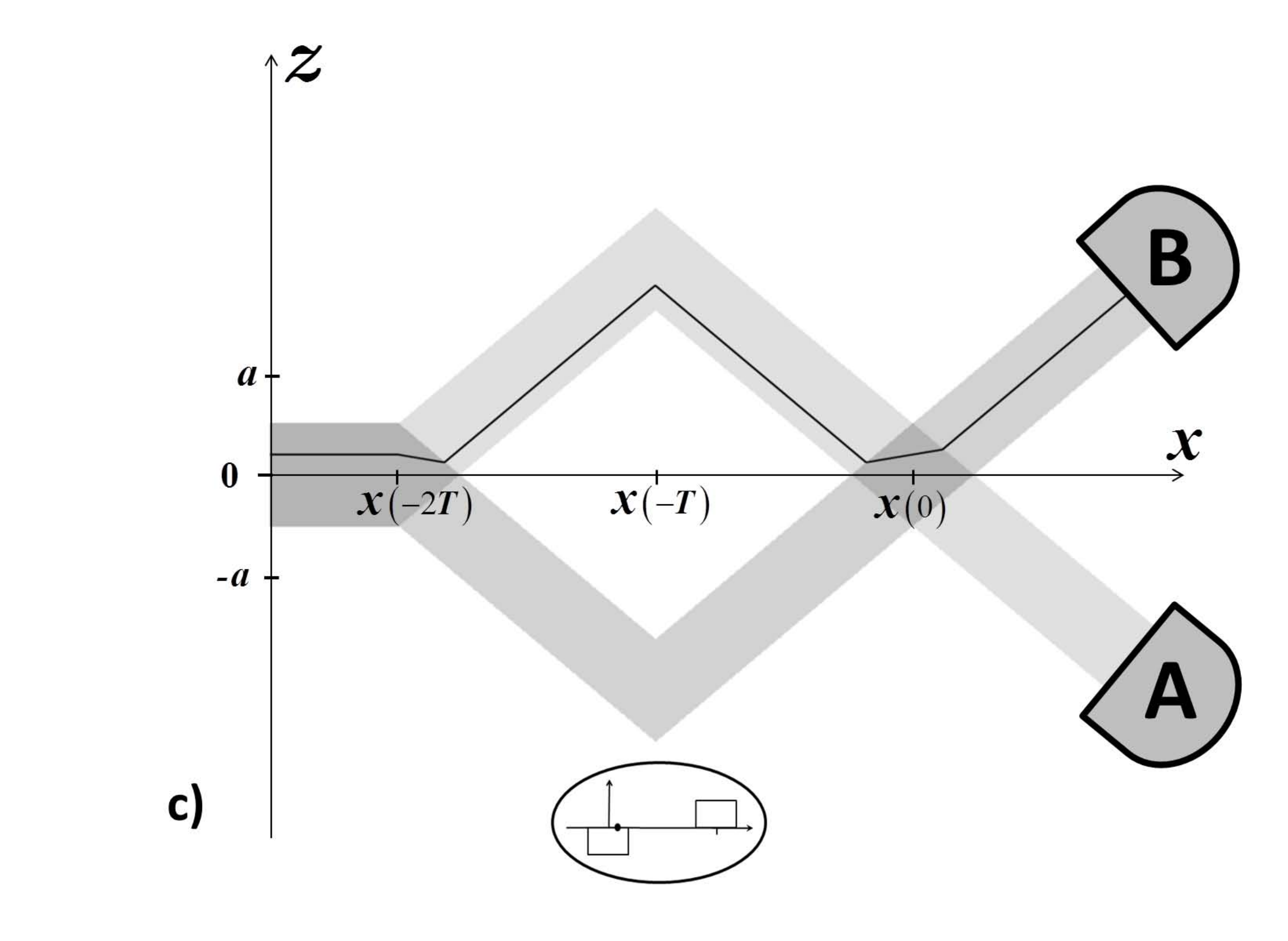}\\ \vspace{-8pt}
     \includegraphics[width=6.3cm]{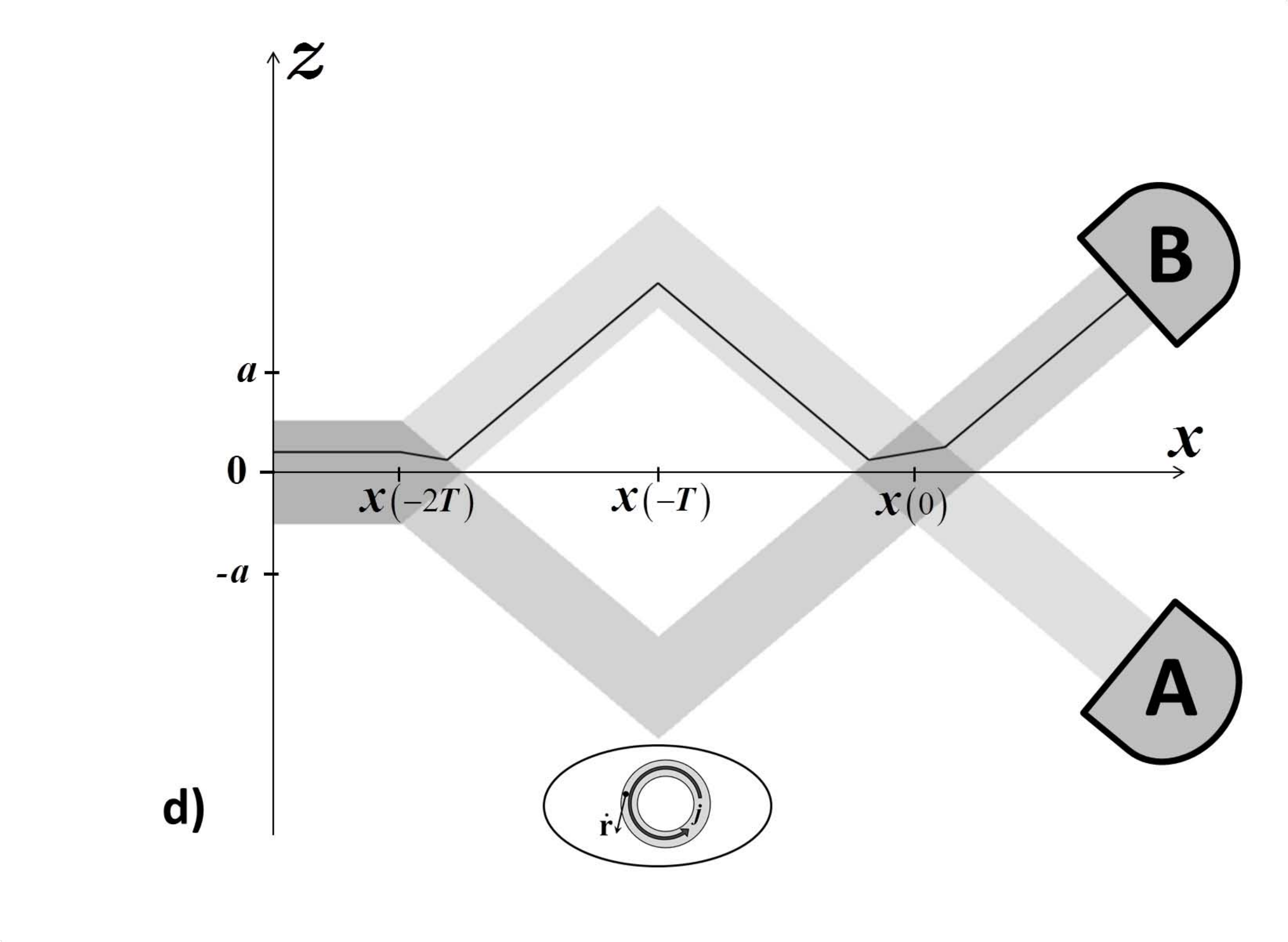}\\ \vspace{-15pt}
    \caption{Trajectories of the particles with various which path detectors.
     a). The Bohmian position detector.  The detector correctly shows that the particle did not pass through the lower arm of the interferometric device.
      b). The spin detector.  The spin flips in spite of the fact that the Bohmian particle did not pass through the lower arm of the interferometric device.
       c). Phase  detector.  The phase changes in spite of the fact that the Bohmian particle did not pass through the lower arm of the interferometric device.
        d). The Bohmian velocity  detector.  The Bohmian velocity changes its direction in spite of the fact that the Bohmian particle did not pass through the lower arm of the interferometric device.
     }
\end{figure}

We can place numerous spin detectors along the arms of the interferometer, and
when we look at them later, they will all show a continuous trajectory
along lower arm which is \textit{not} the Bohmian trajectory. This
phenomenon has received the name ``surrealistic trajectory'' \cite{englert1992surrealistic}.

We have followed Bohm's original prescription for incorporating spin
into the dBB theory. It is, however, sometimes treated differently \cite{DewdneyHKV},
 so it is important to see if the phenomenon of surrealistic trajectories
arises also without spin detectors. We will show that measuring position
with ``phase detector'' (iii) leads to the same phenomenon. The
wave function of the particle and the neutron of the phase detector
for $t>-T$ is:
 \begin{equation}
\frac{1}{\sqrt{2}}\left(\alpha e^{-ik(z+\frac{u}{2}t)}\chi(z+ut)\Big(\chi(\tilde{z}-b)+\chi(\tilde{z})\Big )|\uparrow\rangle+\beta e^{ik(z-\frac{u}{2}t)}\chi(z-ut)\Big(\chi(\tilde{z}-b)-\chi(\tilde{z})\Big )|\downarrow\rangle\right),\label{psioutnox3}
\end{equation}
 The conditional wave function of the particle is either
  \begin{equation}
\alpha e^{-ik(z+\frac{u}{2}t)}\chi(z+ut)|\uparrow\rangle+\beta e^{ik(z-\frac{u}{2}t)}\chi(z-ut)|\downarrow\rangle,\label{psioutnox3+}
\end{equation}
 when the Bohmian position of the neutron is far away from the lower arm of the interferometer, or
  \begin{equation}
\alpha e^{-ik(z+\frac{u}{2}t)}\chi(z+ut)|\uparrow\rangle-\beta e^{ik(z-\frac{u}{2}t)}\chi(z-ut)|\downarrow\rangle,\label{psioutnox3-}
\end{equation}
 when the Bohmian position of the neutron is near the lower arm. In both cases
the Bohmian trajectory of the particle (which started with ${\rm z_{0}}=\frac{a}{5}$)
is in the upper arm  until the overlap with the empty wave, where it turns
toward detector $B$, Fig. 7c. Indeed, the wave function (\ref{psioutnox3+})
is the same as (\ref{psioutnox1}) and the change of sign in (\ref{psioutnox3-}),
does not influence Bohmian trajectory of the particle due to entanglement
with the spin. The particle will be found in detector $B$ and this
will lead to the collapse of the neutron conditional
wave function  to $ \frac{1}{\sqrt{2}}\Big(\chi(\tilde{z}-b)-\chi(\tilde{z})\Big)$
which corresponds to detection of the particle. We again get a surrealistic
trajectory: the phase detector ``detects'' the particle where
the Bohmian position is absent.

Finally, consider a detector of type (iv). It is, in a sense, a Bohmian
detector, since the outcome is written (also) in the behavior of Bohmian
position of the neutron. However, the measurement outcome is registered
in the Bohmian velocity rather than position. The wave function of
the particle and the neutron, for $t>-T$ is:
\begin{equation}
\frac{1}{\sqrt{2\pi R}}\left(\alpha e^{-ik(z+\frac{u}{2}t)}\chi(z+ut)e^{-i\tilde{k}R(\tilde{\theta}+\frac{\omega}{2}t)}|\uparrow\rangle+\beta e^{ik(z-\frac{u}{2}t)}\chi(z-ut)e^{i\tilde{k}R(\tilde{\theta}-\frac{\omega}{2}t)}|\downarrow\rangle\right).\label{psioutnox4}
\end{equation}
 The Bohmian position of the particle (which started with ${\rm z_{0}}=\frac{a}{5}$)
takes the upper arm. Thus, the conditional wave function for the neutron
before $t=0$ is $\frac{1}{\sqrt{2\pi R}}e^{-i\tilde{k}R(\tilde{\theta}+\frac{\omega}{2}t)}$.
Nothing happens at time $t=-T$ to the neutron, its Bohmian position
continue to move on the ring in the clockwise direction since its
conditional wave function does not change. This ``no change''
is, in fact, equivalent to the  detection of the particle in the upper arm, since it is
given that the particle is  in one of the arms. This ``record''
however, does not cause the collapse of the conditional wave function
of the particle as with detector (i): whatever the Bohmian position
of the neutron is, it does not distinguish between rotating wave functions
in different directions, as the probability to find the neutron anywhere
on the ring is constant in both cases. The conditional spatial wave function
of the particle  vary only
in the relative phase which is irrelevant because of the entanglement
with the spin state of the particle. So, once again, formula (\ref{v})
yields that the empty wave packet moving in the lower arm, after the overlap with the wave packet from the upper arm,
will ``capture'' the Bohmian particle and will lead it towards detector $B$, see Fig. 7d.

When the Bohmian position of the particle leaves the overlap area of the wave packets    the conditional wave function of the neutron collapses to
 $\frac{1}{\sqrt{2\pi R}}e^{i\tilde{k}R(\tilde{\theta}-\frac{\omega}{2}t)}$.
Together with this change of the wave function, the rotation of the Bohmian position of the neutron changes from
clockwise to counterclockwise. So, in the end,
we are left with the detector naively showing that the particle passed through the lower arm,
while the Bohmian trajectory passed through the upper arm. The trajectory is surrealistic
again.

Note that the record on detector at lower arm appeared after the
time of the interaction between the particle and the detector. It
is of interest to consider placing detector of type (iv) in the upper arm, instead. A straightforward analysis shows that in this case,
at the time of the interaction between the particle and the neutron,
the neutron will ``record'' the presence of the particle there by changing its conditional wave function and changing the direction
of the velocity of its Bohmian position. However, these records will subsequently
be completely erased, leaving us with ``no change'' in the detector
which corresponds to the particle  passing through
lower arm, demonstrating surrealistic trajectories once again.

\subsection{ Spinless particle}

In all the examples above, the simple formula (\ref{v}) for the velocity
of Bohmian position was exact. This happened because the overlapping
wave packets were entangled with orthogonal spin states and therefore would not interfere with each other. Let us now
see what happens when spin is not involved. For simplicity, we will
consider the same system as above, but flip the spin of the
wave packet moving in the upper arm when it leaves the ``mirror'' which provides the spin-dependent kick, see Fig. 8,
  such that we will have the same spin for both wave packets. The spin will not play any role and we will omit it from our equations.
So, the wave function of the particle for $t>-T$ is
 \begin{equation}
\alpha e^{-ik(z+\frac{u}{2}t)}\chi(z+ut)+\beta e^{ik(z-\frac{u}{2}t)}\chi(z-ut).\label{psioutnospin}
\end{equation}

\begin{figure}[H]\vspace{-10pt}
  \includegraphics[width=10cm]{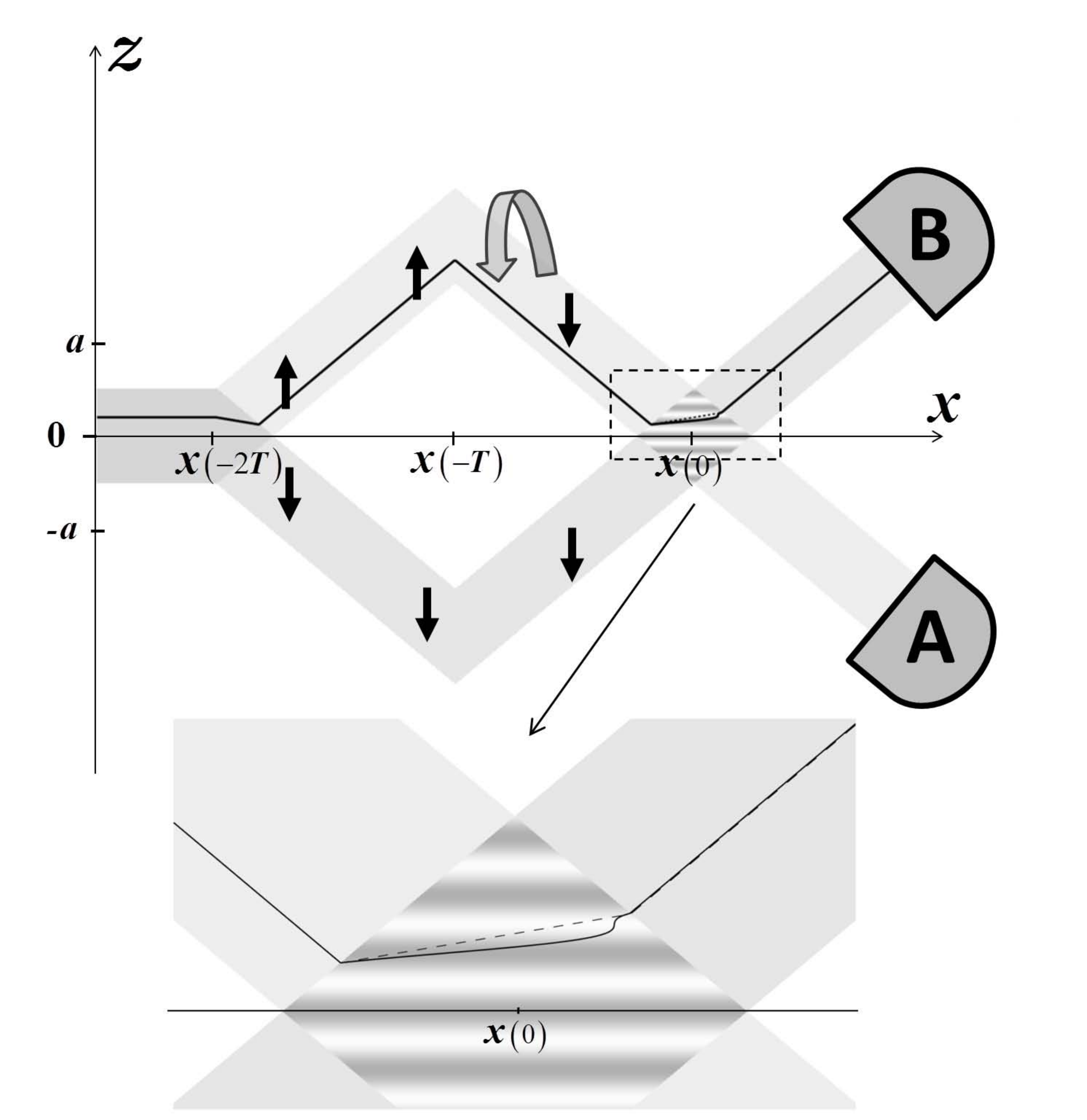}\\\vspace{-5pt}
    \caption{ By flipping the spin in the upper arm, we make it identical for both wave packets and particle's subsequent evolution becomes as if it had no spin. The amplitude of the wave function $|\psi(x,z,\frac{x-{\rm x}_0}{v})|$ (the shade of the grey color) shows  the interference pattern in the overlap area. It exhibits discontinuity on the boundary of the overlap area which is an artifact of our simplified model. In this model the Bohmian trajectory  $\mathrm{z}(x)$ (continuous line) slightly deviates from  $\mathrm{z}_{appr}(x)$ (dashed line), the trajectory calculated on the basis of approximate formula (\ref{v}). }
\end{figure}

 Now formula (\ref{v})  does not hold exactly. We have to use the
basic equation (\ref{rdot}). The difference appears in the area of
the overlap where the two wave packets interfere, see Fig. 8, which shows interference for $\lambda=\frac{2\pi}{k}=0.3a$. In contrast with
(\ref{v}), the calculations of exact Bohmian trajectory are more
complicated, but for the simple idealized case we consider here, we
can get an implicit analytic solution. In the area of the overlap,
(\ref{rdot}) yields for Bohmian velocity
\begin{equation}
\dot{\mathrm{z}}=\frac{(|\beta|^{2}-|\alpha|^{2})u}{1+2|\alpha\beta|\cos(2k\mathrm{z}+\phi)},\label{zdotns}
\end{equation}
 where $\quad\frac{\alpha|\beta|}{|\alpha|\beta}=e^{-i\phi}$. Integrating,
we get an implicit solution:
\begin{equation}
\mathrm{z}=(|\beta|^{2}-|\alpha|^{2})ut-\frac{|\alpha\beta|}{k}\sin(2k\mathrm{z}+\phi)+C_{1}.\label{zns}
\end{equation}
 The solution based on formula (\ref{v}) is
  \begin{equation}
\mathrm{z}_{appr}=(|\beta|^{2}-|\alpha|^{2})ut+C_{2}.\label{zns1}
\end{equation}
 The constants are specified by the continuity of $\mathrm{z}(t)$
at the boundary of the overlap region. The equations limit the difference
between the constants $|C_{1}-C_{2}|\leq\frac{|\alpha\beta|}{k}$
and thus the actual Bohmian position oscillates near approximate solution.
In our example, $\alpha$ and $\beta$ are real, so $\phi=0$. The
initial conditions for the Bohmian particle position are such that
the Bohmian particle enters the overlap region at time $t=-\frac{3a}{8u}$,
when $\mathrm{z}=\frac{a}{8}$. These parameters define the constants:
$C_{1}=\frac{a}{8} +\frac{\sqrt 6}{5k}\sin\frac{2ka}{5}$ and $C_{2}=\frac{a}{8}$.

After an integer number of oscillations, $\mathrm{z}(t)=\mathrm{z}_{appr}(t)$, so, if the Bohmian particle leaves the overlap
region at that moment,  it will continue
to move exactly as described by formula (\ref{v}).  In fact, this is a general feature which is
always valid \cite{Gold}. Our calculations, however, show that the approximate and exact trajectories  slightly differ when they leave the overlap area.
 The reason for the discrepancy is our usage of   an inconsistent model of non-spreading wave packet with sharp edges.
The  inconsistency of the model   is transparent when we
realize  the discontinuity of the wave function  at the boundary of the overlap area.

 Considering the exact evolution of the wave packets
eliminates the discontinuity of the wave function. Indeed, the deviation disappears
when we consider a Gaussian model, in Sec. 4.3.
 Since in the $x$ direction the Bohmian particle moves with constant
velocity, $\mathrm{z}(t)$ essentially gives the trajectory $\mathrm{z}(x)$.
This exact trajectory makes small oscillations (of the order of one tens of the wavelength) about the trajectory
calculated using formula (\ref{v}) and coincides with it when the
wave packets are separated.

Let us consider now what happens when we add a position detector.
We will start again with detector (i) which records the outcome on
the Bohmian position of a neutron.
We put our detector in the lower arm of the device. The wave
function of the particle and the neutron for $t>-T$ is
 \begin{equation}
\alpha e^{-ik(z+\frac{u}{2}t)}\chi(z+ut)\chi(\tilde{z})+\beta e^{ik(z-\frac{u}{2}t)}\chi(z-ut)\chi(\tilde{z}-b).\label{psioutnospin2}
\end{equation}
 Consider again the initial Bohmian position of the particle ${\rm z_{0}}=\frac{a}{5}$.
Since at time $t=-T$ the Bohmian trajectory of the particle  is in the upper arm, the conditional wave function of the neutron
remains $\chi(\tilde{z})$ and thus, its Bohmian position remains
unchanged: ${\rm \tilde{z}}\in[-\frac{a}{2},\frac{a}{2}]$. Then,
at time $t=0$ the conditional wave function of the particle is:
 \begin{equation}
e^{-ik(z+\frac{u}{2}t)}\chi(z+ut).\label{psiout5}
\end{equation}
 Therefore, the Bohmian position of the particle will continue to
ride down along with wave packet moving towards detector $A$, at the location $\frac{a}{4}$
below center of the wave packet. We see that with or without
the spin, the Bohmian trajectory is the naively expected straight line, Fig. 7a.

Let us consider a spin detector of type (ii). The measured particle
has no spin (it can be mimicked by a particle with spin that does not
change), but the neutron has spin. In fact, that is essentially all it has; we
assume that its spatial state is not changed in the process of measurement.
The wave function of the particle and the neutron for $t>-T$ is:
\begin{equation}
\alpha e^{-ik(z+\frac{u}{2}t)}\chi(z+ut)|\tilde{\uparrow}\rangle)+\beta e^{ik(z-\frac{u}{2}t)}\chi(z-ut)|\tilde{\downarrow}\rangle).\label{psioutnox33ns}\end{equation}
 The orthogonal spin states of the neutron entangled with the wave
packets of the particle play the same role as the spin of the particle
in the analysis above. As in the case of a particle with  spin, we get a surrealistic trajectory with a straight
line segment corresponding exactly to the formula (\ref{v}), see Fig. 7b.

Our next case is the phase detector (iii), see Fig. 9a. The wave function of the
particle and the neutron of the phase detector for $t>-T$ is
 \begin{equation}
\frac{1}{\sqrt{2}}\left(\alpha e^{-ik(z+\frac{u}{2}t)}\chi(z+ut)\Big(\chi(\tilde{z}-b+\chi(\tilde{z}))\Big)+\beta e^{ik(z-\frac{u}{2}t)}\chi(z-ut)\Big(\chi(\tilde{z}-b)-\chi(\tilde{z})\Big)\right).\label{psioutnox3ns}
\end{equation}
 The conditional wave function of the particle is either
  \begin{equation}
\alpha e^{-ik(z+\frac{u}{2}t)}\chi(z+ut)+\beta e^{ik(z-\frac{u}{2}t)}\chi(z-ut),\label{psioutnox3+ns}
\end{equation}
 when the Bohmian position of the neutron $\tilde{\mathrm{z}}$ is near $b$, or
 \begin{equation}
\alpha e^{-ik(z+\frac{u}{2}t)}\chi(z+ut)-\beta e^{ik(z-\frac{u}{2}t)}\chi(z-ut),\label{psioutnox3-ns}
\end{equation}
 when the Bohmian position of the neutron is near the origin. In the first
case the conditional wave function is identical to the wave function
of the particle without position detector (\ref{psioutnospin}),
so $\mathrm{z}(t)$ is given by (\ref{zns}) and the trajectory is identical to the one shown in Fig. 8.
In the second case (\ref{psioutnox3-ns}), the detector causes
a minor modification of the trajectory due to the change in the phase of  the oscillations,
in (\ref{zns}) $\phi\rightarrow\phi+\pi$, see Fig. 9a. So, in both cases we get surrealistic
trajectories with small oscillations around the solution obtained on the basis of formula (\ref{v}).

Finally, consider detector of type (iv). The wave function of the
particle and the neutron, for $t>-T$ is:
 \begin{equation}
\frac{1}{\sqrt{2\pi R}}\left(\alpha e^{-ik(z+\frac{u}{2}t)}\chi(z+ut)e^{-i\tilde{k}R(\tilde{\theta}+\frac{\omega}{2}t)}+\beta e^{ik(z-\frac{u}{2}t)}\chi(z-ut)e^{i\tilde{k}R(\tilde{\theta}-\frac{\omega}{2}t)}\right).\label{psioutnox4ns}
\end{equation}
\begin{figure}[H]\vspace{-10pt}
     \includegraphics[width=8cm]{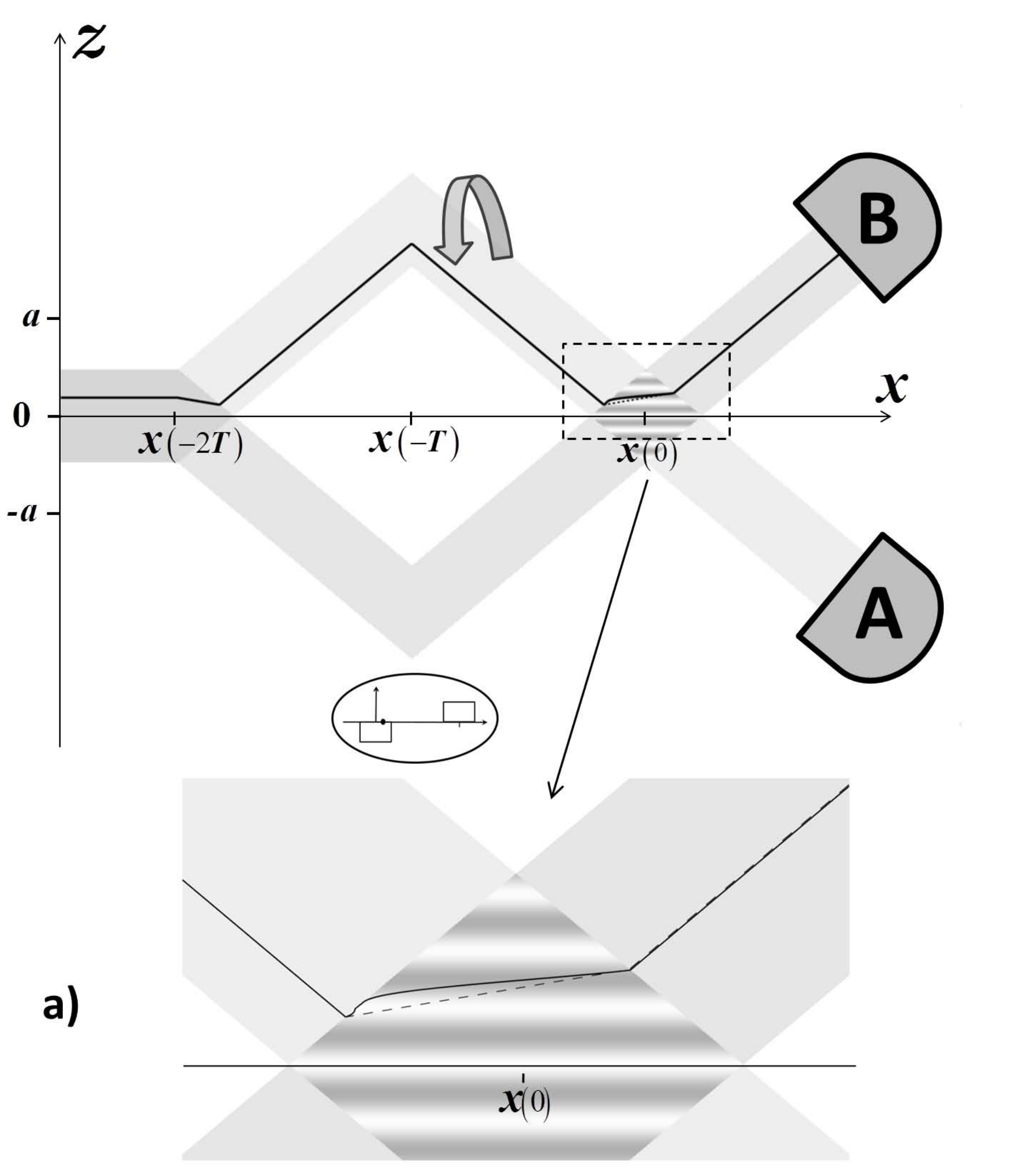}\\\vspace{-5pt}
         \includegraphics[width=8cm]{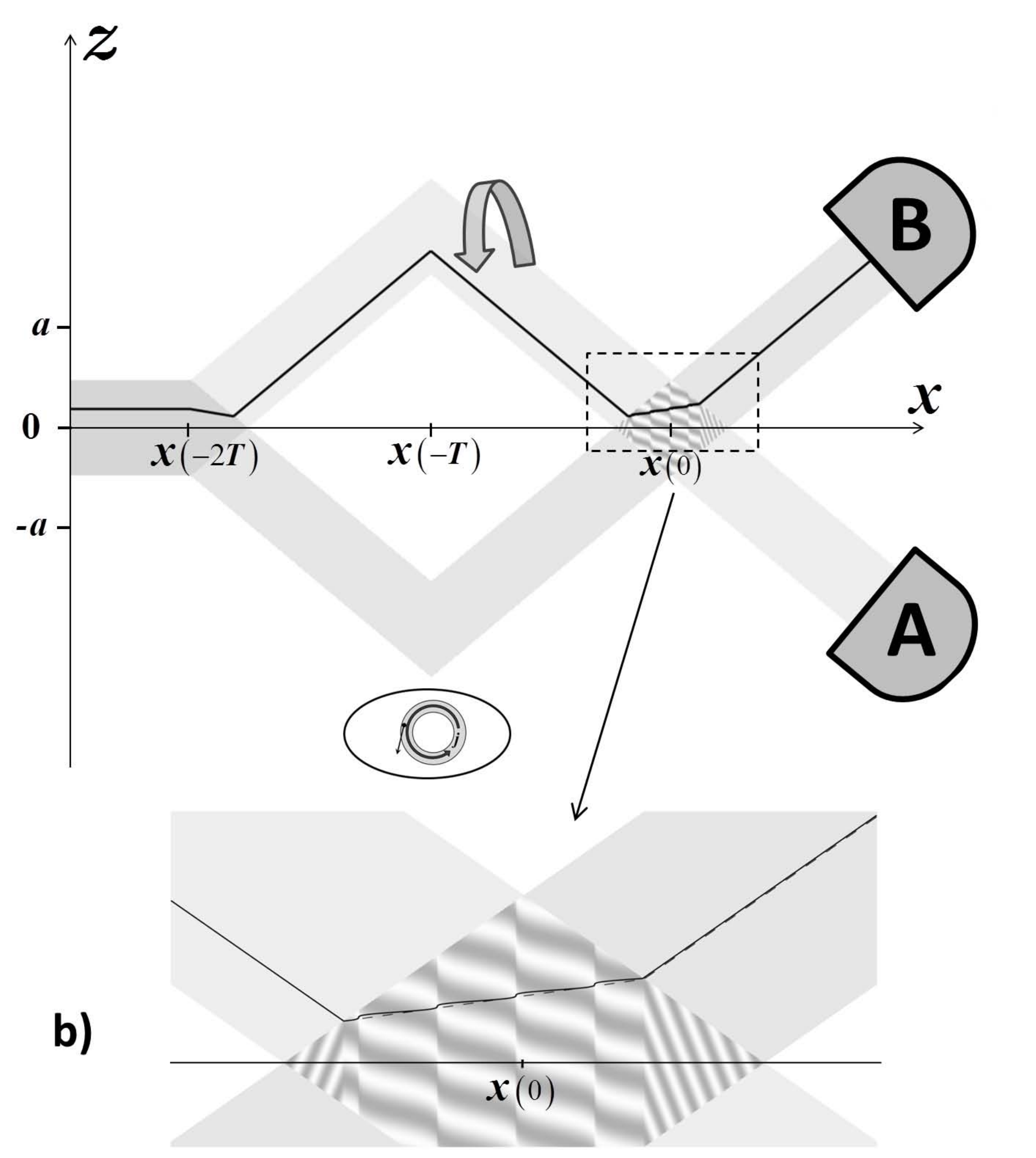}\\\vspace{-10pt}
    \caption{a). Bohmian  trajectory of a spinless particle when a ``phase detector'' placed at the lower arm clicked.
   b). Bohmian  trajectory of a spinless particle when a Bohmian velocity  detector placed at the lower arm clicked.
   For comparison the  trajectory calculated on the basis of formula (\ref{v}) is shown by a dashed line.}
\end{figure}
If,  as in our case, the Bohmian particle moves in the upper arm,
the Bohmian position of the neutron continues to move clockwise, while
if the Bohmian particle moves in the lower arm, then at time $t=-T$, the
neutron switches its velocity to counterclockwise direction.

When the wave packets of the particle overlap, at time $t=T$, both,
the Bohmian positions of the particle and of that of the neutron, move. For
finding the analytic solution it is helpful to consider this as a
motion of a Bohmian position of a fictitious ``particle'' in two dimensions, $z,\tilde{\theta}$.
We assume that the mass of the neutron equal to the mass of the particle and change the basis to a ``rotated basis''
 \begin{eqnarray}
Z & = & \frac{kz+\tilde{k}R\tilde{\theta}}{\sqrt{k^{2}+\tilde{k^{2}}}},\nonumber \\
 & ~\\
Y & = & \frac{-\omega Rz+uR\tilde{\theta}}{\sqrt{u^{2}+\omega^{2}R^{2}}},\nonumber
\end{eqnarray}
 such that at the region of the overlap the wave function becomes
a superposition of two plane waves moving in the $\hat{Z}$ and $-\hat{Z}$
directions,
\begin{equation}
\frac{1}{\sqrt{2\pi R\epsilon a}}\left(\alpha e^{-i\sqrt{k^{2}+\tilde{k^{2}}}\left(Z+\frac{u}{2}\sqrt{1+\frac{\tilde{k^{2}}}{k^{2}}}t\right)}+
\beta e^{i\sqrt{k^{2}+\tilde{k^{2}}}\left(Z-\frac{u}{2}\sqrt{1+\frac{\tilde{k^{2}}}{k^{2}}}t\right )}\right).\label{Zofx}\end{equation}
 Thus, in the overlap, the Bohmian trajectory will have constant $Y$
and oscillations around straight line in ${\rm Z}(x)$ of the amplitude
$\frac{|\alpha\beta|}{\sqrt{k^{2}+\tilde{k^{2}}}}$.

Returning back to the conditional wave function of the particle inside the interferometer, we see that the detector causes a significant change of the interference picture, see Fig. 9b. It is obtained for $\tilde{\lambda}=\frac{2\pi}{\tilde{k}}=0.1a$.  Thus, the trajectory is even
closer to the one predicted by the formula (\ref{v}) than the trajectory without the detector. The oscillations
are more frequent, but the amplitude of the oscillations, $\frac{|\alpha\beta|k}{ \tilde{k}^{2}+ k^{2}}$, is  smaller by another order of magnitude.

\subsection{ Gausssian wave packet}

In all the examples above we were able to make exact analytical calculations,
either directly with formula (\ref{v}) or with the exact formula
(\ref{rdot}). In all cases, formula (\ref{v}) gave either the exact,
or a very good approximation to the exact trajectory. However, we considered
only a simple model with a rectangular wave function neglecting its
spread, which is clearly inconsistent. (We have seen one aspect of
the inconsistency above.) In order to test the behavior of the Bohmian
trajectories we will avoid sharp edges in the wave function and will
take the spreading of the wave function into account. We replace $\chi(z)$,
the wave function of the particle at time $-2T$, by the Gaussian
wave function $\frac{1}{a\sqrt{\pi}}e^{-\frac{z^{2}}{2a^{2}}}$. The wave function
of the particle for $t>-T$ is
\begin{equation}
{\cal N}(t)\big(\alpha e^{-ik(z+\frac{u}{2}t)}e^{-\frac{(z+ut)^{2}}{2[a^{2}+i(t+2T)\frac{u}{k}]}}+\beta e^{ik(z-\frac{u}{2}t)}e^{-\frac{(z-ut)^{2}}{2[a^{2}+i(t+2T)\frac{u}{k}]}}\big).\label{gaus}
\end{equation}

\begin{figure}[H]\vspace{-20pt}
  \includegraphics[width=9cm]{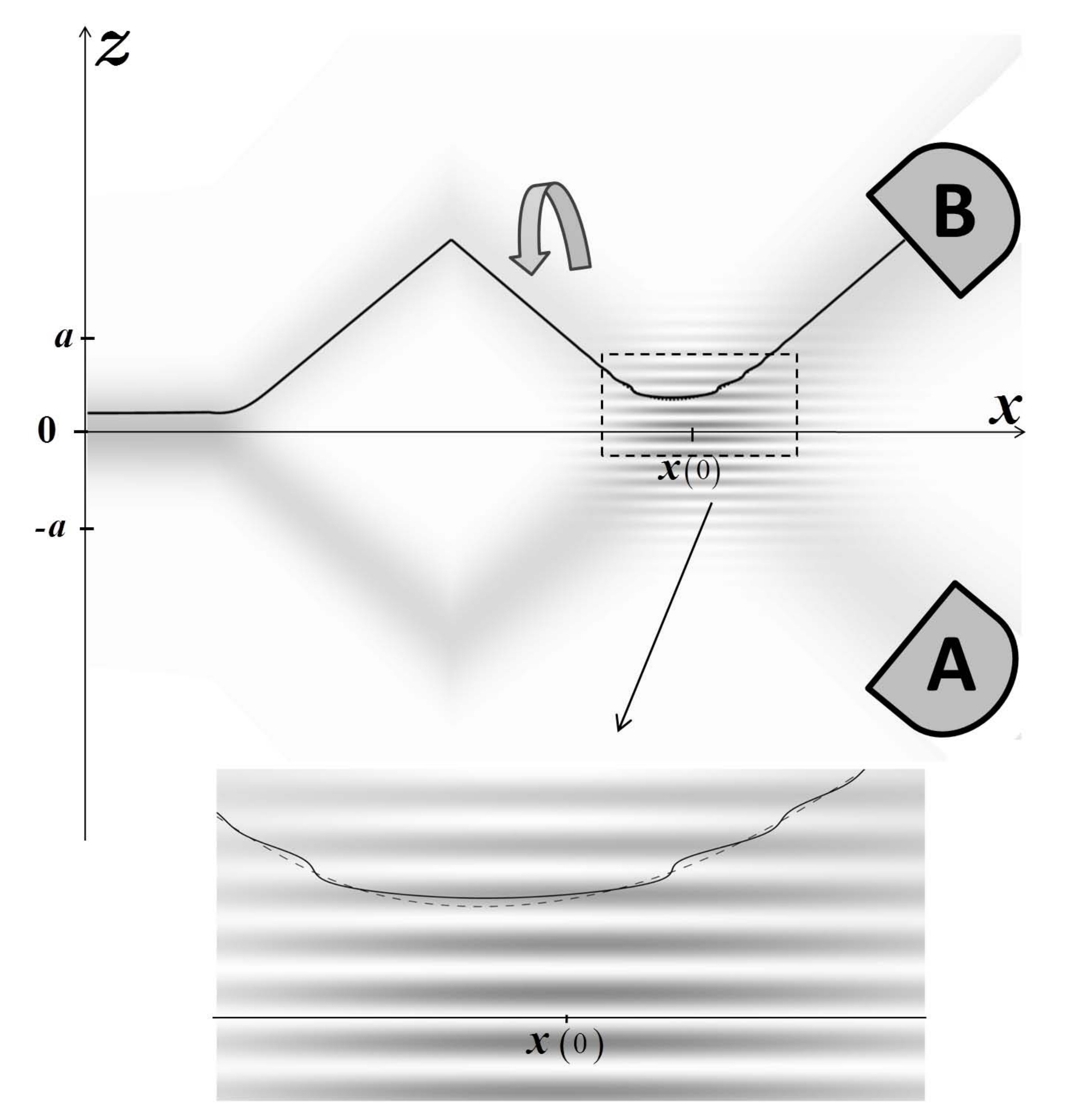}\\\vspace{-9pt}
      \caption{ Bohmian trajectory of a  spinless particle for a Gaussian wave packet. Exact trajectory (solid line) compared with formula (\ref{v}) calculations (dashed line).}
\end{figure}

\vspace{-10pt}
 We will consider the initial Bohmian position ${\rm z}_{0}=0.37a$.
It is similar to the condition ${\rm z}_{0}=\frac{a}{5}$ for the
rectangular wave function (\ref{chi1}) since in both cases $70\%$
of the weight of the wave packet is below ${\rm z}_{0}$. The results
of numerical calculations are shown in Fig.10. First we see that the formula (\ref{v}) yields good approximation at the overlap area and exact solution outside the overlap area.
We see also that qualitative
behavior of the trajectory is the same as with rectangular wave packet
shown in Fig.8, the Bohmian particle position ``changes hands''
at the overlap region between the wave packet moving at the upper arm and the wave packet moving towards detector $B$. However,
some of the features we do not see: we cannot recognize a straight
line in the overlap regions with the slope $|\beta|^{2}-|\alpha|^{2}$.
The explanation is that  the trajectory passes through regions
far away from the centers of the Gaussians, where the derivative of
the amplitude is relatively large in contrast with vanishing derivative for  rectangular wave packets. At the center of the Gaussian,
the derivative is zero, so we can expect a better correspondence.
To test this let us consider the same wave function, but with initial
 Bohmian positions at the center ${\rm z}=0$. Then, it
will also pass through the center of the overlap region where the
derivatives of the amplitudes of both Gaussian wave packets vanish.
Figs. 11a, 11b, clearly show the similarity of the trajectories for rectangular
and Gaussian wave packets.  In both cases the trajectory performs small oscillations
around the solution based on (\ref{v}). Our numerical solution of the exact equations of motion
indicates the applicability  of the simplified model of rectangular wave
packets used throughout the paper. Moreover, we can see that for Gaussian wave function the  trajectory
coincides with the solution based on (\ref{v}) when  the wave packets separate, contrary to the case of  rectangular wave
packets where they  slightly deviate.

\begin{figure}[H]\vspace{-10pt}
  \includegraphics[width=8cm]{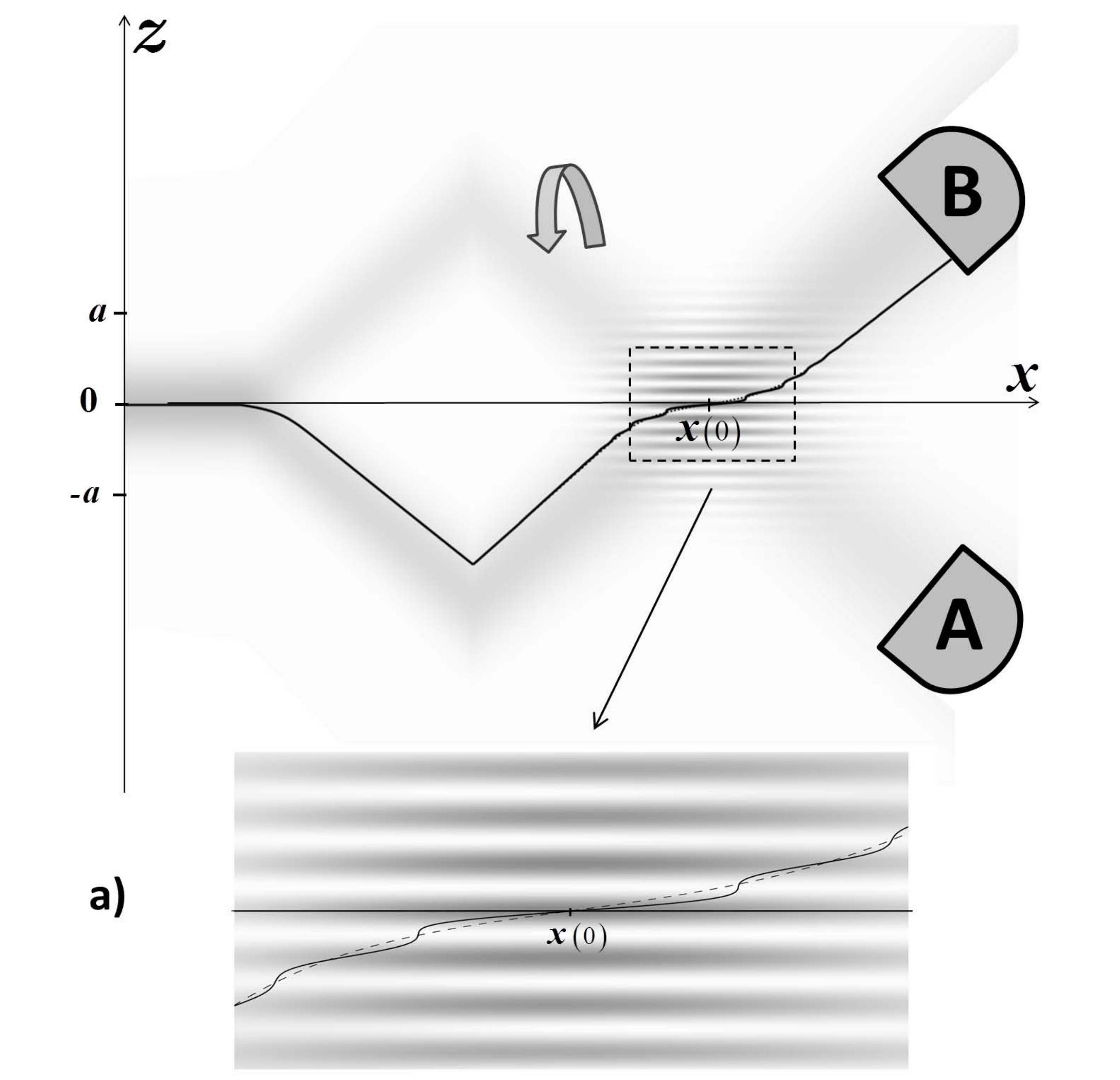}\\\vspace{-10pt}
  \includegraphics[width=8cm]{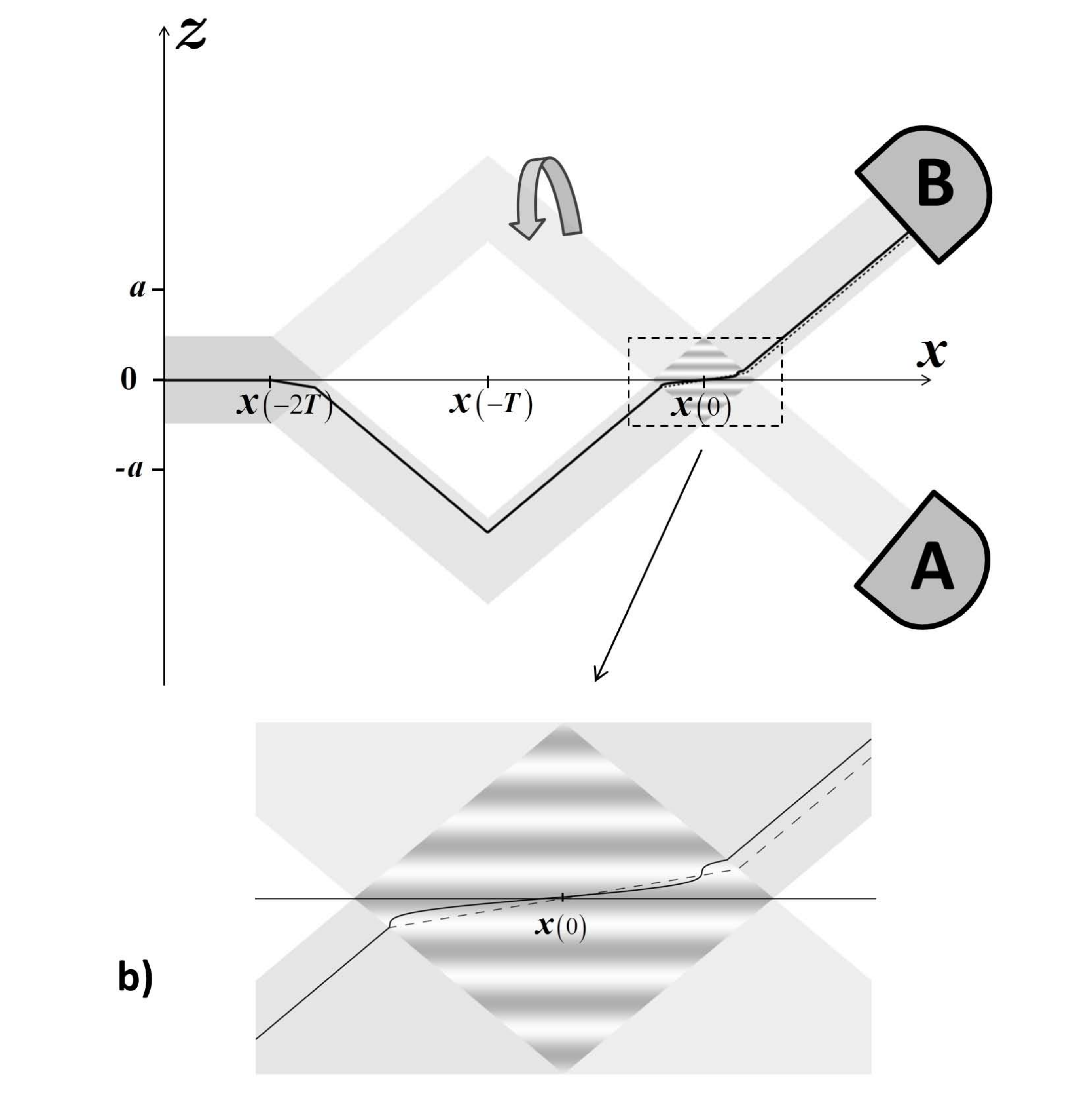}\\\vspace{-10pt}
    \caption{a). Bohmian trajectory passing through the center of the overlapping region. a) Exact trajectory for Gaussian wave packet (solid line) compared with formula (\ref{v}) calculations (dotted line). b). Trajectory calculated according to exact local equation but under the assumption of nonspreading rectangular wave function (solid line) compared with formula (\ref{v}) calculations (dashed line).}
\end{figure}

\section{ Discussion and conclusions}

It was not our aim here to argue in favor or against the dBB theory
in comparison with other interpretations as was done by Tumulka \cite{tumulka:1220}. In our view, due to tremendous difficulties
with collapse and bizarre reality of the many-worlds interpretation,
the dBB theory is also a  legitimate theory for developing intuition in quantum interference
experiments.   This picture might be useful even if we
do not accept the ontological claims of the dBB theory. Note recent experimental demonstration of average trajectories in a two-slit interferometer \cite{Steinberg2011}.

 The main message
of our paper is that the simple formula for Bohmian velocities (\ref{v})
is widely applicable for analysis of Bohmian mechanics. In most cases the formula is
either exact, or provides a very good approximation. It allowed us to analyze
the behavior of various position detectors uncovering a nontrivial
structure in ``surrealistic trajectories''.  We can
put aside the question of applicability of no crossing
theorem and just make simple explicit   estimation of Bohmian trajectories.
There are numerous basic quantum experiments, like tunneling and scattering off a barrier
which are very difficult to understand with the intuition we have
from observing the classical world. The proposed approximate, yet very simple, Bohmian picture
is very helpful for the analysis of these quantum phenomena.

We would also like to add a comment on the criticism   by Hiley and Callaghan \cite{hiley2006delayed} of ``slow bubble
chamber model'' \cite{kill} proposed by one of the authors a few years ago.
The ``slow bubble chamber model'' is a hybrid of detectors of
the types (i) and (iv). The wave packet of the neutron gets a momentum
kick if the particle is present as in (iv), but its wave packet is
not a constant wave on the ring, but a relatively wide free wave packet.
The kick is weak, so it will take a long time until the kicked and
unkicked wave packets separate as in (i). If the experiment ends before
this separation, we get surrealistic trajectories since the slow bubble
chamber works as a detector of type (iv). In the conceptual model
considered by Vaidman \cite{kill} everything in the ``bubble'' was slow
(contrary to a real bubble chamber in which electron ionization happens
fast). We can strengthen Vaidman's argument  by using a SG splitting
instead of the usual beam splitter, since then the formula (\ref{v}) becomes exact.
Note, however, that the usual beam splitter also demonstrates well the argument of
this paper due to approximate validity of  (\ref{v}).

\section*{Acknowledgements}

It is a pleasure to acknowledge the Towler Institute for organizing
the workshop  ``21st-Century directions in de Broglie-Bohm theory
and beyond''. The authors LV and GN benefited tremendously from discussions
at this meeting. This work has been supported in part by the Binational
Science Foundation Grant No. 32/08, the Israel Science Foundation
Grant No. 1125/10.

\end{document}